\documentclass[pre,eqsecnum,showpacs,amssymb]{revtex4}
\usepackage{amsmath,graphicx}
\usepackage{epsfig,subfigure}
\usepackage{bm}

\begin{document}

\title{Forced translocation of a polymer: dynamical scaling vs. MD-simulation}

\author{J. L. A. Dubbeldam$^{2}$, V. G. Rostiashvili$^1$, A.
Milchev$^{1,3}$, and T.A. Vilgis$^1$}
\affiliation{$^1$ Max Planck Institute for Polymer Research, 10 Ackermannweg,
55128 Mainz, Germany\\
$^2$ Delft University of Technology 2628CD Delft, The Netherlands\\
$^3$ Institute for Physical Chemistry, Bulgarian Academy of Sciences, 1113
Sofia, Bulgaria}

\begin{abstract}
We suggest a theoretical description of the force-induced  translocation
dynamics of a polymer chain through a nanopore. Our consideration is based on
the tensile (Pincus) blob picture of a pulled chain and the notion of
propagating front of tensile force along the chain backbone, suggested recently
by T.~Sakaue. The driving force is associated with a chemical potential
gradient that acts on each chain segment inside the pore. Depending on its
strength, different regimes of polymer motion (named after the typical chain
conformation, ``trumpet'', ``stem-trumpet'', etc.) occur. Assuming that the
local driving and drag forces are equal (i.e., in a quasi-static approximation),
 we derive an equation of motion for the tensile front position $X(t)$. We show
that the scaling law for the average translocation time $\langle \tau \rangle$
changes from $ \langle \tau \rangle \sim N^{2\nu}/f^{1/\nu}$ to $ \langle \tau
\rangle \sim N^{1+\nu}/f$ (for the free-draining case) as the dimensionless 
force ${\widetilde f}_{R} = a N^{\nu}f /T$ (where $a$ , $N$ , $\nu$, $f$, $T$ 
are the Kuhn segment length, the
chain length, the Flory exponent, the driving force, and the temperature,
respectively) increases. These and other predictions are tested by Molecular
Dynamics (MD) simulation. Data from our computer experiment indicates indeed
that the translocation scaling exponent $\alpha$ grows with the pulling force
${\widetilde f}_{R}$) albeit the observed exponent $\alpha$ stays systematically
smaller than the theoretically predicted value. This might be associated with
fluctuations which are neglected in the quasi-static approximation.
\end{abstract}


\maketitle

\section{Introduction}

The force-induced translocation through a pore in the membrane is mainly
motivated by the possibility for fast DNA and RNA sequencing
\cite{Zwolak,Branton}. Translocation dynamics is an essential component of
transport in biological cells \cite{Van_der_Laan}. Most of the experimental
studies so far deal with a driven polymer translocation which is realized by
applying an electrical field across a narrow pore \cite{Aksimentiev}. However,
despite extensive research, numerous computer experiments, and a variety of
attempts for theoretical interpretations, the translocation dynamics even in the
generic case remains at present insufficiently well understood
\cite{Lubensky,Muthu,Chern,Klafter,Grosberg,Kantor,Dubbeldam_1,Panja}. As
emphasized in a recent review paper \cite{Milchev}, currently there exists  a
plethora of theoretical predictions for the value of the translocation exponent
$\alpha$ which governs the scaling of the mean translocation time $\langle \tau
\rangle$ with the length $N$ of the polymer chain  ($\langle \tau \rangle
\propto N^{\alpha}$). These values vary in a rather broad interval: one predicts
$\alpha = 1$ \cite{Lubensky,Muthu,Klafter}, or $\alpha = 2\nu + 1 - \gamma_1
\approx 1.49$ \cite{Dubbeldam_1} for a three-dimensional ($3D$) self-avoiding
(SAW) polymer chain (where the Flory exponent $\nu \approx 0.59 $ and  the
surface exponent $\gamma_1 \approx 0.68$), up to $\alpha = (1+2\nu)/(1+\nu)
\approx 1.37$ \cite{Panja}  and $\alpha = 1+\nu \approx 1.59 $
\cite{Kantor,Grosberg}.

In a recent paper \cite{Dubbeldam} we demonstrated that the dynamics of a single
polymer moving through a nanopore in a membrane (i.e. the translocation
dynamics) with or without an external force being present can be treated within
the framework of fractional Brownian motion (fBm). It was shown that the
corresponding Fokker-Planck equation of motion (FPEM) contains time-dependent
drift and diffusion terms. In the case of non-driven chain this FPEM leads
naturally to anomalous diffusion of the translocation coordinate  $s(t)$
\cite{Dubbeldam} that describes the number of segments threaded through the pore
at time $t$. On the other hand, for a driven translocation fluctuations play a
relatively moderate role and one can directly construct reasonable scaling
relations that describe the process. Recently, an interesting approach based on
the notion of tensile force propagation along the chain (within the well known
picture of tensile blobs\cite{Brochard_1, Brochard_2}) has been suggested by T.
Sakaue \cite{Sakaue_1,Sakaue_2} in an effort to provide a consistent description
of translocation dynamics. Sakaue's approach appears physically sound and well
capable of providing a plausible interpretation of existing observation.
However, his mathematical treatment requires the use of a ``cut-off'' trick as 
applied to the polymer segments density function (see more details in 
the Appendix \ref{App}). This trick is  questionable and affects the
predicted translocation exponent $\alpha$. 

In the present paper we follow Sakaue's approach and consider theoretically the
case of a driven translocation by means of a different derivation which does not
require additional conjectures. In Sec. II we investigate the various regimes
(named traditionally a ``trumpet'', ``stem-trumpet'', and ``stem'' regimes),
observed for weak, intermediate and strong driving forces. Our consideration is
based on equating the driving and drag forces, i.e., on a {\em quasi-static
approximation} whereby fluctuations in this balance of forces are ignored. Our
main concern is the scaling law for the mean translocation time $\langle \tau
\rangle$ with respect to the chain length $N$ and the applied driving force $f$.
We find $\langle \tau \rangle \propto N^{\alpha} f^{-\delta}$ albeit with a
different expression for $\alpha$ as well as  the mean translocated length
$M(t)$ vs. time.

We also perform extensive Molecular Dynamics (MD) simulations in order to test
our theoretical predictions. In Sec. III we describe the simulation model and
compare theoretical predictions with simulation data. This comparison shows that
the exponent $\alpha$, found in MD-simulation, is systematically smaller then
its theoretical estimate even though $\alpha$ changes with $f$ as expected. We
argue that this may be due to fluctuations which are neglected in the
theoretical treatment. Finally, in the Appendix \ref{App} we explain in detail
why the calculations of Sakaue \cite{Sakaue_1,Sakaue_2} are bound to produce a
scaling exponent that differs from the proper one. 

\section{Dynamical scaling picture of a driven translocation}
\label{sec:theory}

\subsection{A general overview}

We assume that the pulling  force is applied only to a bead which is inside the
pore, i.e., the potential drops mainly across the pore. This assumption was
recently examined for charged DNA translocations \cite{Wanunu,Grosberg_1}.  It
was shown, in particular, that the ionic current generates a nonuniform electric
field which is extended far beyond the pore and acts on both the DNA  and its
counter ions. In this case the DNA threading through the pore is preceded by its
capture by electroosmotic flow. On the other hand, it can be shown that for a
sufficiently narrow pore the voltage drop mainly takes place inside the pore
\cite{Kowalczyk}. In this situation the translocation is mainly determined by
the threading process which is the main subject of our investigation.

When a pulling force is suddenly switched on, tension starts to propagate along
the chain backbone which alters the polymer conformation progressively.
Eventually, after some characteristic time a steady state is reached and the
whole polymer starts moving with constant velocity. Such a non-equilibrium
response is of great importance in many biological and technological situations.
Below we follow to some extent the model developed in the recent papers by
Sakaue \cite{Sakaue_1,Sakaue_2}.
\begin{figure}[ht]
\includegraphics[scale=0.34]{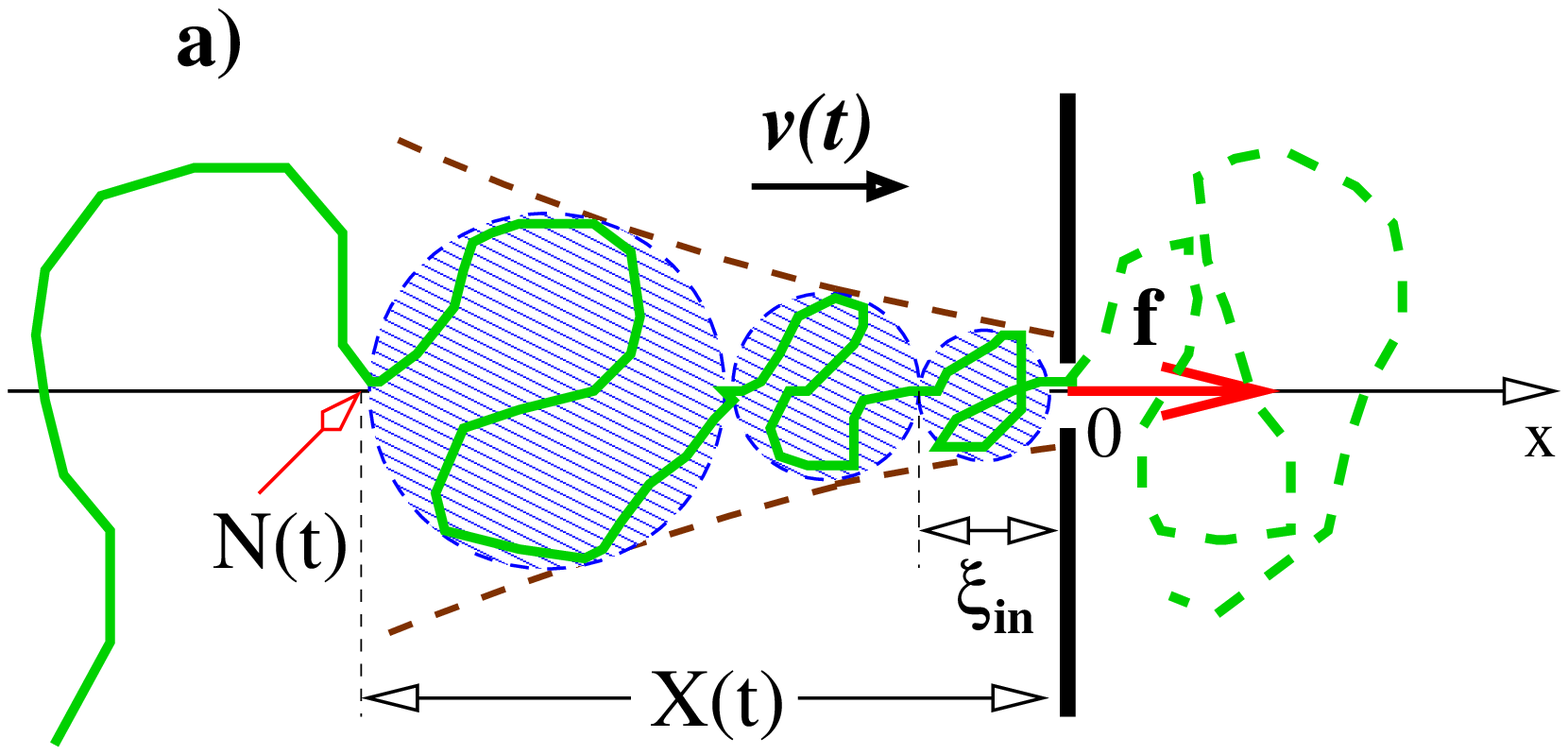}
\hspace{0.5cm}
\includegraphics[scale=0.34]{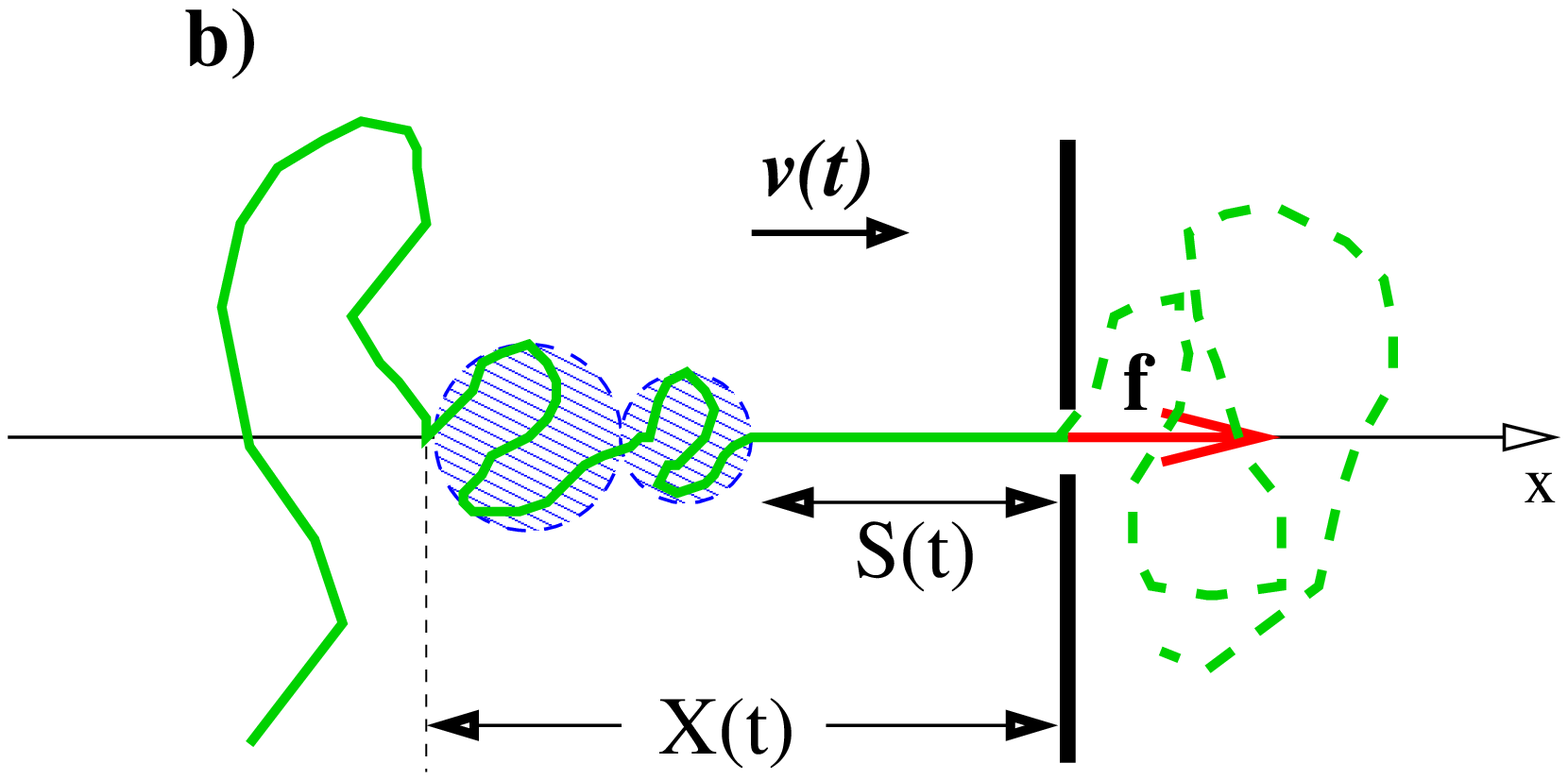}
\hspace{0.5cm}
\includegraphics[scale=0.31]{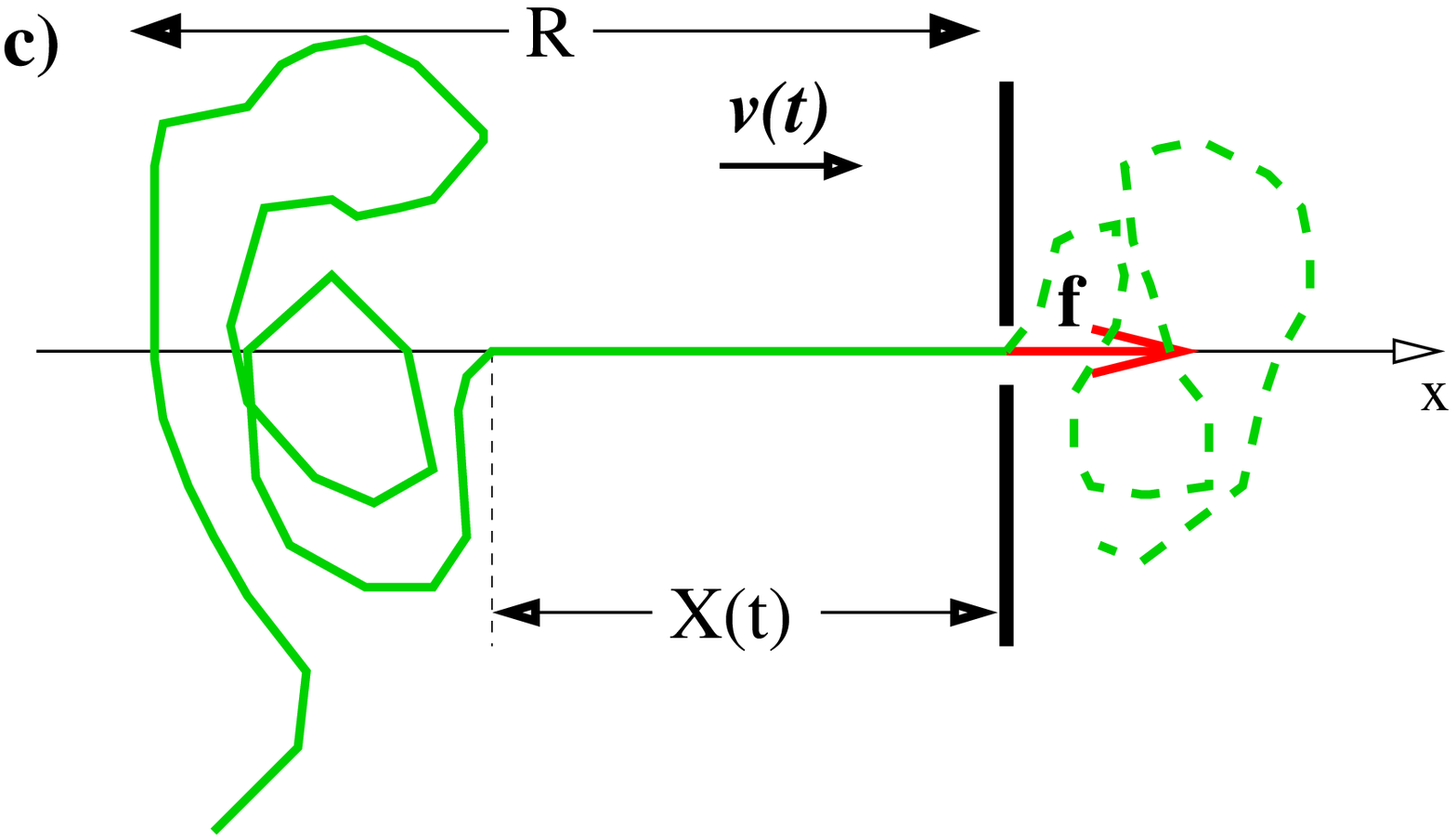}
\caption{(Color online) Dynamical response of the polymer chain shortly after the driving force
$f$ has been switched on. By the time $t$ the tension is transmitted up to the
$N(t)$-th monomer. The number of the translocated monomers is denoted by $M(t)$
(indicated by a dashed line on the $trans$-side of the pore). The distance
between the propagating tension front and the membrane is marked as $X(t)$. The
portion of the chain marked as $X(t)$ is moving with an average time-dependent
velocity $v(t)$. a) The ``trumpet'' regime at driving force $k_B T/aN^{\nu} \ll
f \leqslant k_BT/a$ . The initial blob size is $\xi_{\rm in} = k_B T/f$. b) At
$k_BT/a \leqslant f \ll (k_BT/a) N^{\nu}$ the part of the chain affected by
tension starts as a ``stem'' of the length $S(t)$ and then turns into a
``trumpet'' (``stem-trumpet'' regime). c) At $f > (k_BT/a) N^{\nu}$ the tensed
part of the chain is completely stretched (``stem'' regime). $R$ marks the
unperturbed chain size, i.e., $R = a N^{\nu}$.} \label{Transient}
\end{figure}

Figure \ref{Transient} shows  schematically the case of {\it tensile force
transmission}. The driving as well as the friction forces are $x$-dependent
which leads for a relatively small driving force exerted on a segment in the
pore, $k_B T/aN^{\nu} \ll f \leqslant k_BT/a$, to the so-called  ``trumpet''
blob picture (shaded area in Fig.~\ref{Transient}a)
\cite{Brochard_1,Brochard_2}. By equating the local driving and drag forces (see
Eq. (\ref{Balance})) and using the relationship between the tensile blob size
$\xi(x)$ and the force $f(x)$ (cf. Eq. (\ref{Pincus_Blob})), one could derive
the ``trumpet'' profile equation (cf. Eq. (\ref{Solution_Blob}) ). The location
of the tension front $X(t)$ is defined by the free boundary condition $f(x = -
X(t)) = 0$. This condition along with the material balance Eq.
(\ref{Material_Balance}) and the closure relation, Eq. (\ref{Closure}), leads to
the differential equation, Eq. (\ref{Transmission_Diff}), for the front
propagation  $X(t)$. The characteristic time of this transmission is given as
$\tau_1 = C_1 \: N^{1+\nu}/f^{z-1 - 1/\nu}$ (cf. Eq. (\ref{Tau_1})), where the
dynamical exponent $z = 2+1/\nu$ for the free-draining (or Rouse) case, and $z =
3$ for the non-draining (or Zimm) case, $C_1$ being a model dependent constant.
For a Rouse chain, $\tau_1 = C_1 \: N^{1+\nu}/f$  which agrees with the scaling
law predicted by Kantor and Kardar \cite{Kantor}. During the process of tensile
force transmission (or, front propagation), the velocity $v(t)$ of the moving
domain decreases (because as time goes by more and more segments get
incorporated in the moving domain) and at the time $t = \tau_1$ it approaches a
stationary value $v_s$. After that a stationary regime sets in and the rest of
the chain is sucked into the pore with a constant velocity $v_s$ (cf. Eq.
(\ref{Velocity_Stationar})). The characteristic time of this {\it stationary
suction} process is $\tau_2 = C_2 N^{2\nu}/f^{z-2}$, (see Eq. (\ref{Tau_2})). The
 times $\tau_1$ and $\tau_2$ add up to a net translocation time $\langle \tau
\rangle = \tau_1 + \tau_2$. Generally, the force transmission time $\tau_1$
prevails for a strong enough driving force and a long chain (see Eq.
(\ref{Condition})). Thus, the translocation time scaling ranges between
$N^{2\nu}/f^{z-2}$ and $N^{1+\nu}/f^{z-1 - 1/\nu}$, depending on the pulling
force intensity.

For moderate, $k_B T/a  \leqslant f \ll (k_B T/a)  N^{\nu}$, and strong, $(k_B
T/a) N^{\nu} < f$, forces, the ``trumpet'' regime is replaced by the
``stem-trumpet'', see Fig.~\ref{Transient}b, and ``stem'', see
Fig.~\ref{Transient}c, regimes respectively. In the ``stem-trumpet'' case, the
part marked as $S(t)$ of the total  moving domain looks like a stem whereas the
rest resembles a trumpet shape. For the ``stem'' scenario of the force
transmission the total moving domain looks like a completely stretched portion,
see Fig.~\ref{Transient}c. In both regimes the characteristic times of the force
transmission and stationary suction into the pore are $\tau_1 \sim N^{1+\nu}/f$,
and $\tau_2 \sim N^{2\nu}/f$. It could be seen that the time $\tau_1$ dominates
for a longer chain (see Eq. (\ref{Long_Chain})). In the Table \ref{tab:Label} we
summarize the results of the theoretical scaling prediction for the
translocation time $\langle \tau \rangle \propto N^{\alpha} f^{-\delta}$ (for
the Rouse case) as well as the corresponding MD-simulation findings which will
be discussed in this Sec. \ref{sec:theory} and in Sec. \ref{Numerics}.
\begin{table}[htbp] 
\begin{tabular}{|c|c|c|c|}
 \hline
Regime & Exponents $\alpha$ and $\delta$ (theory)& Conditions & Exponents
$\alpha$ and $\delta$  (MD)\\ \hline
  Trumpet:  & \quad $\alpha = 2\nu \approx 1.18$ ,\quad $\delta = 1/\nu \approx
1.66$
 \quad &  ${\widetilde f}_R \ll (C_2/C_1)^{\nu/(1 - \nu)}$ & $\alpha \approx
1.11$, \quad $\delta \approx 1.17$ \\ &  $\alpha = 1+\nu \approx 1.59$, $\delta
= 1$ & ${\widetilde f}_R \gg (C_2/C_1)^{\nu/(1 - \nu)}$ & $\alpha \approx 1.47$
,
\quad $\delta \approx 0.97 $ \\
 & see Eq. (\ref{Tau_Translocation})& see Eq.(\ref{Condition})& see Sec.
\ref{Numerics}\\ \hline
Stem-Trumpet & $\alpha = 2\nu\approx 1.18 $ , \quad $\delta = 1$ & short chain &
\\
and Stem: & $\alpha = 1 + \nu \approx 1.59$ ,  $\delta = 1$ & longer chain & \\
& see Eq. (\ref{Transloc_Time_Stem_Coil})& see Eq.(\ref{Long_Chain}) & \\
\hline
\end{tabular}
\caption{Exponents $\alpha$ and $\delta$ for the translocation time scaling
$\langle \tau \rangle \propto N^{\alpha} f^{-\delta}$ in the free-draining
(Rouse) case as predicted by the theory (theory) and by our MD-simulation (MD)
(see Sec. \ref{Numerics}). The parameter ${\widetilde f}_R \stackrel{\rm def}{=}
a N^{\nu} f/k_B T$, whereas $C_1$ and $C_2$ are some numerical model-dependent
parameters.} \label{tab:Label}
\end{table}

\begin{figure}[ht]
\includegraphics[scale=0.3]{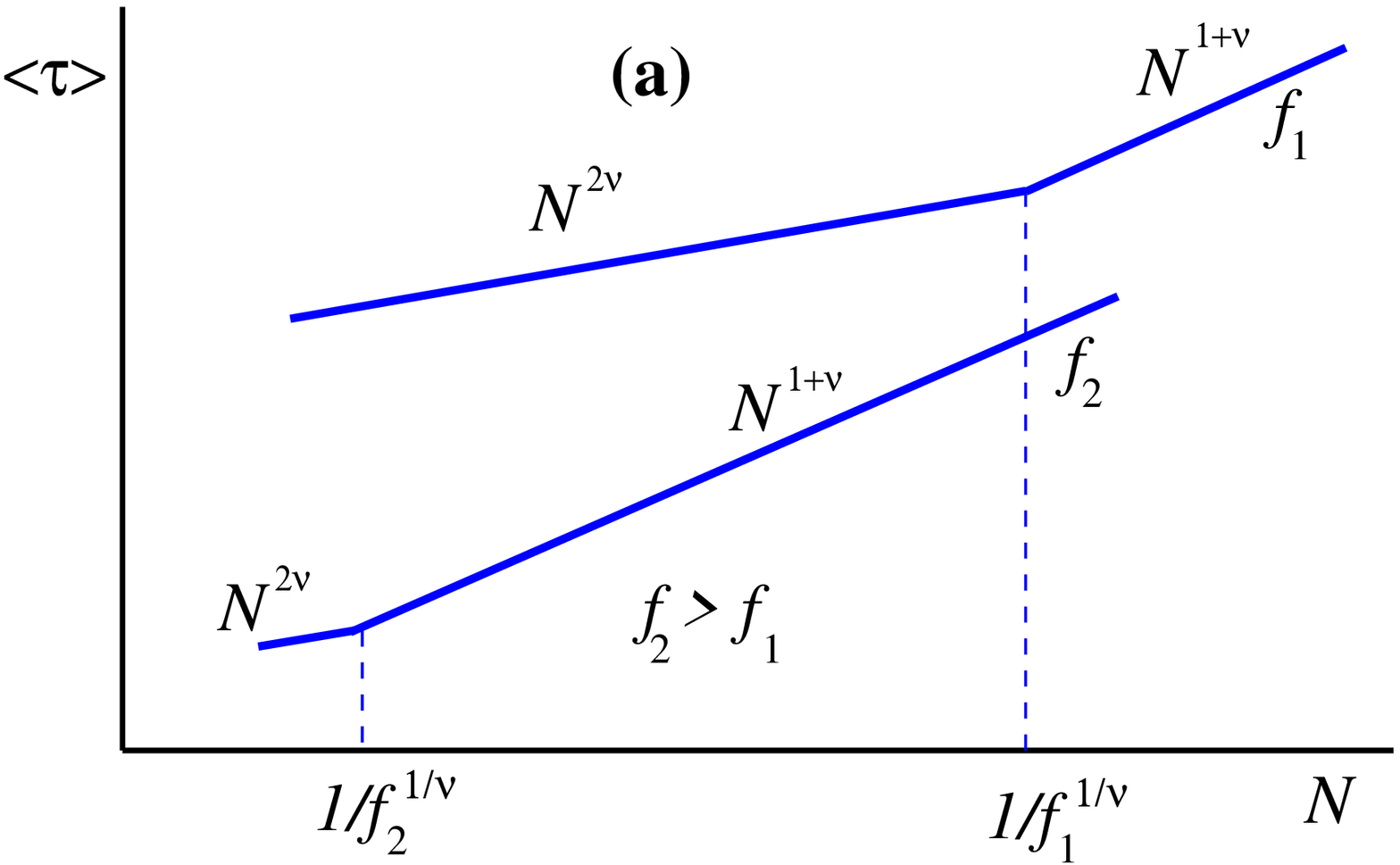} 
%\vspace{1cm}
\includegraphics[scale=0.3]{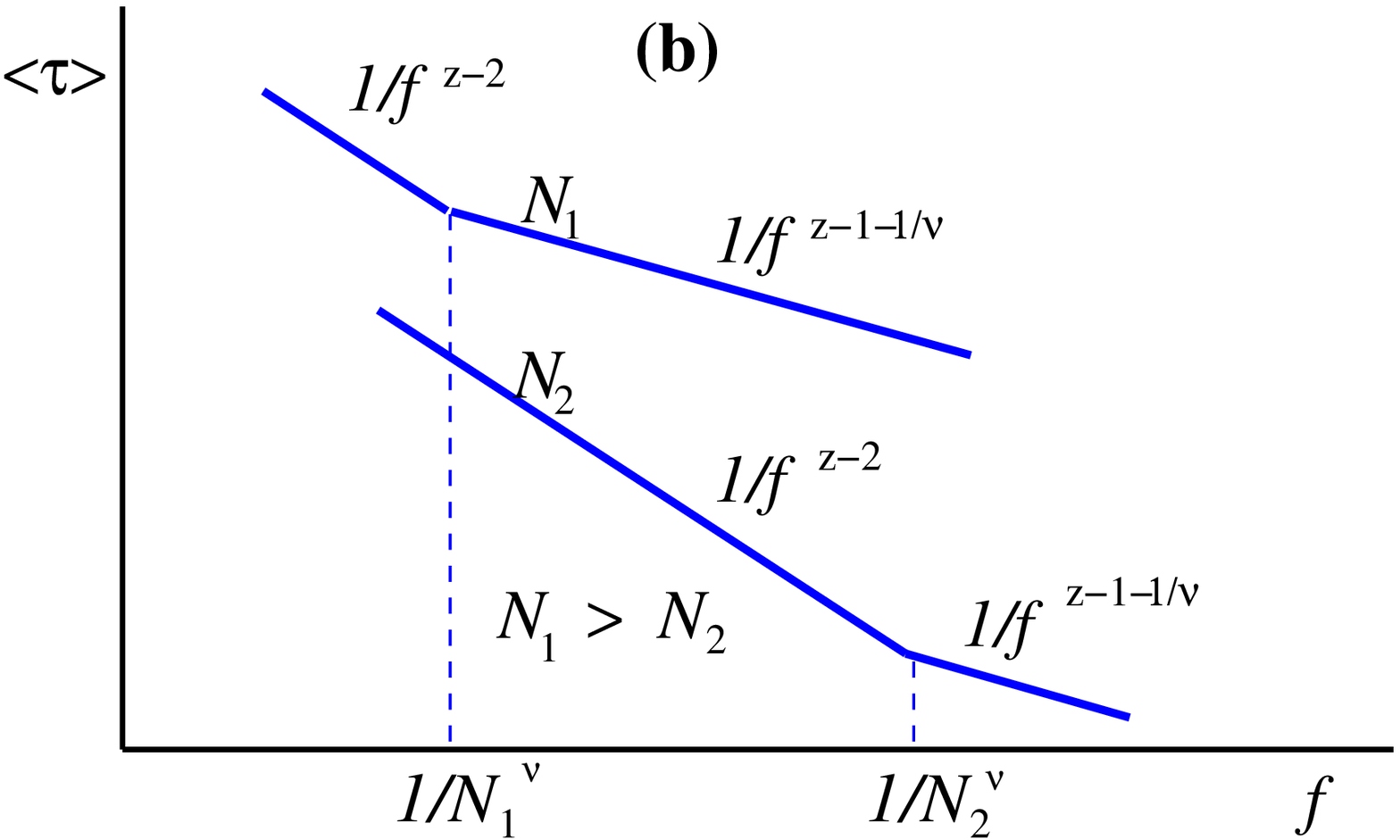} 
%\vspace{1cm}
\includegraphics[scale=0.3]{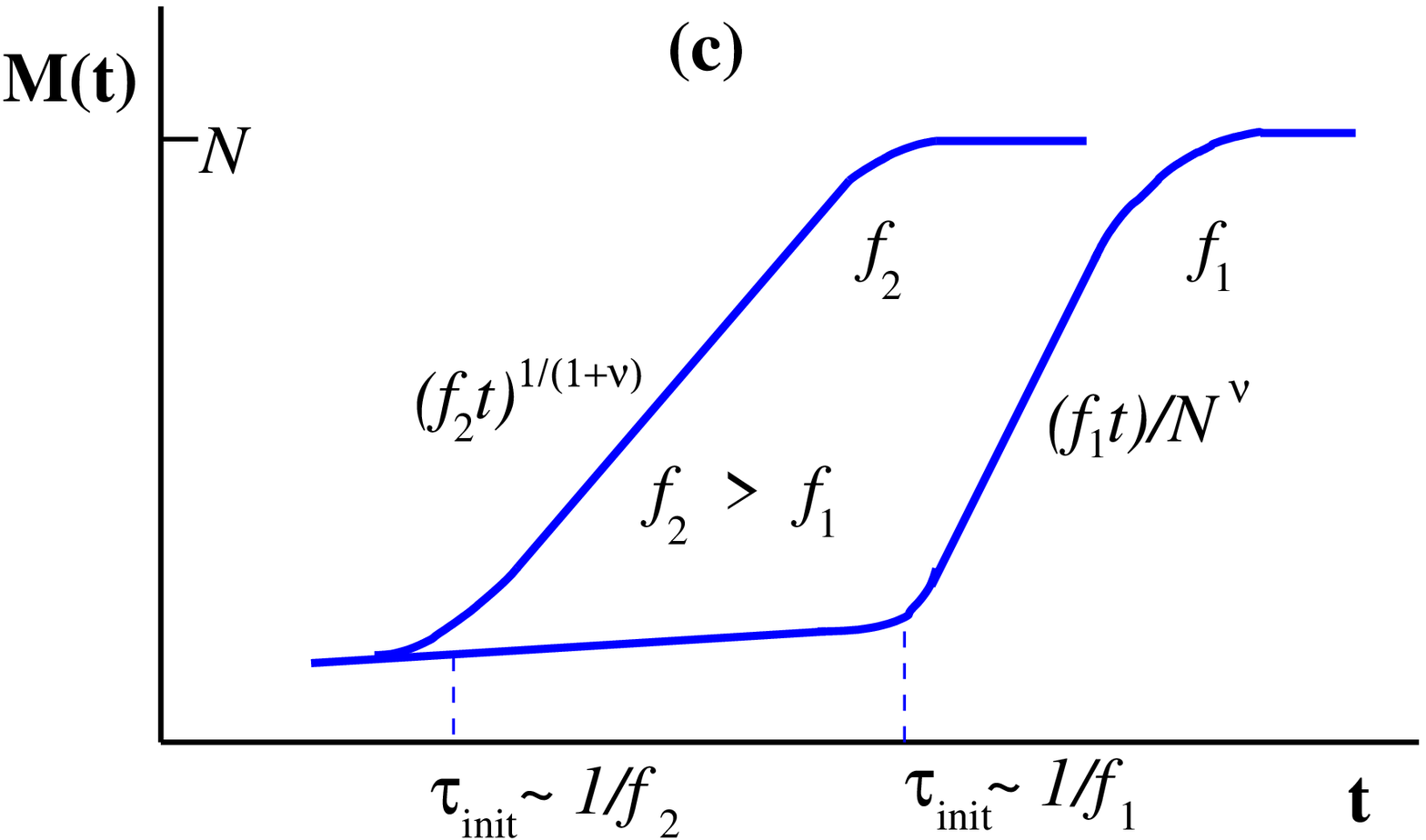} 
\caption{ Schematic representation of the main scaling predictions: a)
Translocation time $\langle \tau\rangle$ vs. chain length $N$ at a fixed
force $f$. With growing force ${\widetilde f}_R \stackrel{\rm def}{=} a
N^{\nu} f/k_B T$ the scaling law $\langle \tau\rangle \sim N^{\alpha}$ changes
from $\langle \tau\rangle \sim N^{2 \nu}$ to $\langle \tau\rangle \sim N^{1
+ \nu}$. b) Translocation time $\langle \tau\rangle$ vs. driving force $f$ at
different chain lengths $N_1 > N_2$. c) Translocation length $M(t)$ vs. time $t$
for two different forces $f_2 > f_1$. The characteristic time for initial blob
formation goes as $\tau_{\rm init} \sim 1/f$.}
\label{Schematic}
\end{figure}

Schematically, our theoretical findings for the scaling behavior of $\langle
\tau \rangle$ for the ``trumpet'' regime (see Sec. IIC) are plotted in Fig.
\ref{Schematic}a (for $\langle \tau \rangle \sim N^{\alpha}$ ) and in
Fig.\ref{Schematic}b (for  $\langle \tau \rangle \sim f^{- \delta}$). The time
evolution of the number of translocated monomers  $M(t)$ is also investigated in
Sec. IIC2. Again, depending on driving force intensity, the $M(t)$ vs. time  $t$
scaling law changes, see Fig.\ref{Schematic}c, from $M(t) \propto (f
t)^{1/(1+\nu)}$ (at strong force, when the force transmission dominates)  (see
Eq. (\ref{Translocation_Law_Rouse})) to $M(t) \propto (f t)/N^{\nu}$ (for weak
force, when the stationary suction process dominates and $t \ll \tau_2$) (see
Eq. (\ref{Trans_Monomers_Stat_2})). These $M(t)$ vs. $t$ dependencies clearly
illustrate the presence of an initial stage,  which  in the Sec. IIB  is called
{\it blob initiation}, and is confirmed by our MD-results (see Fig.
\ref{fig:savt} where the translocation coordinate $\langle s \rangle
\stackrel{\rm def}{=} M(t)$ vs. time $t$ is given). It will also be shown that
the characteristic time of this initial stage $\tau_{\rm init}$ changes with
force as $\tau_{\rm init} \sim 1/f$ (see Eq. (\ref{Initiation_1}) and Fig.
\ref{fig:fig5}.

We will devote the rest of this Sec. \ref{sec:theory}  to the detailed
discussion of the ``trumpet'' , ``stem-trumpet'' and  ``stem'' regimes as well
as to the mean translocation length $M(t)$ behavior. Readers who are not
interested in the mathematical aspects of the problem can skip these details and
go to Sec. \ref{Numerics}. One should keep in mind, however, that before one or
another type of tensile force transmission comes into play, an initial blob
should always set in.

\subsection{Blob initiation}
\label{BI}

The starting deformation (i.e., {\it blob initiation}) of a polymer chain that
has been initially at rest and is then pulled by a force $f$ constitutes a
necessary stage to set in the driving force transmission. An initial blob of 
size $\xi_{\rm init} = k_BT/f$ is formed which contains $g$ beads so that
$\xi_{\rm init} = a g^{\nu}$ where $a$ is the Kuhn segment length and $\nu$ is
the Flory exponent, mentioned above. Denote $\tau_0 = a^2 \zeta_0/k_BT$ as the
characteristic diffusion time of a single bead with $\zeta_0$ being the Stokes
friction coefficient \cite{deGennes}. Then the blob initiation time can be
written in the scaling form
\begin{eqnarray}
\tau_{\rm init} = \tau_0 \: \phi \left(\dfrac{\xi_{\rm init}}{a}\right)
\label{Initiation}
\end{eqnarray}
so that the scaling function $\phi(y)$ depends only on the dimensionless
combination $\xi_{\rm init}/a$ (i.e., on the size of the blob itself). On the
other hand, we assume that the initial blob formation happens due to purely
mechanical traction which almost does not affect the  chain meanders. This is
supported by the fact that within the time interval $0 < t < \tau_{\rm init}$
only few segments are translocated (see Fig. \ref{Schematic}c). Thus, the
characteristic time $\tau_{\rm init}$ does not depend on temperature $T$ and is
determined only by the Stokes friction coefficient $\zeta_0$ and the pulling
force $f$. Since $\tau_0 \propto 1 / k_BT$, this implies $\phi(y) \propto y$,
and one gets as a result
\begin{eqnarray}
 \tau_{\rm init} = \left( \dfrac{a^2 \zeta_0}{k_BT} \right)\:
\left(\dfrac{k_BT}{a f}\right) \propto \dfrac{\zeta_0 a}{f}
\label{Initiation_1}
\end{eqnarray}
The rearrangement of the initial blob leads at $t > \tau_{\rm init}$ to a
subsequent tensile force transmission which is governed by the local balance of
driving and drag forces. The blob initiation characteristic time $\tau_{\rm
init}$ is indeed clearly seen in our MD-simulation results (see
Fig.\ref{fig:savt}) and found to change as $\sim 1/f$. In the next sections we
consider different regimes of this tensile force transmission, depending of the
strength of $f$, which are named after some characteristic shapes of the
resulting chain conformation.

\subsection{Trumpet regime}
\label{TR}

In this case the driving force is moderately strong and falls within the range
\begin{eqnarray}
 \dfrac{k_BT}{a N^{\nu}} \ll f \leqslant \dfrac{k_BT}{a}
\label{Forces_Range}
\end{eqnarray}
Schematically, a typical conformation of a driven polymer in the trumpet regime
is represented by the shaded area in Fig. \ref{Transient}. At a given time $t$
some fraction of the chain is subjected to tension. This part of the chain
comprises a sequence of Pincus blobs of different size,
\begin{eqnarray}
 \xi (x) = \dfrac{k_BT}{f(x)},
\label{Pincus_Blob}
\end{eqnarray}
where $f(x)$ is the local (i.e. $x$-dependent) value of the driving force. As is
evident from  Fig. \ref{Transient}, the blob size at a given position $x$
corresponds to the lateral chain meandering at $x$. According to the blob
definition $\xi (x) = a g(x)^{\nu}$ where $g(x)$ is the number of beads in the
blob (we recall that the chain inside the blob follows Self Avoiding Walk -
statistics). On the other hand, as stipulated by the local force balance, the
driving force equals the friction (Stokes) force, i.e.
\begin{eqnarray}
 f(x) =  \zeta_0 \: v(t)  \int\limits_{-X(t)}^{x} \:
\left[\dfrac{\xi(x')}{a}\right]^{z-2} \: \dfrac{dx'}{\xi(x')} \label{Balance}
\end{eqnarray}
In Eq.(\ref{Balance}) we take into account that $d x' / \xi(x')$ is the number
of blobs on the interval $\{x', x'+dx'\}$ whereas $\zeta_0 \: v(t) \;
[\xi(x')/a]^{z-2}$ is the local Stokes friction force. The dynamical exponent
$z$ in Eq. (\ref{Balance}), which determines the total number of monomers in
this portion of the chain, is $z = 2 + 1/\nu$ for the case of Rouse dynamics,
and $z = 3$ for the Zimm statistics \cite{deGennes}. Thus, taking into account
Eq.~(\ref{Pincus_Blob}), the expression for the blob size becomes
\begin{eqnarray}
 \xi(x) = \dfrac{k_BT}{\zeta_0 \: v(t) \; \int_{-X(t)}^{x} [\xi(x')/a]^{z-2} d
x'/\xi(x')}\; , \label{Balance_1}
\end{eqnarray}
or, represented in a differential form,
\begin{eqnarray}
 \dfrac{d \xi}{d x} = - \dfrac{\zeta_0 \: v(t)}{k_B T \: a^{z-2}} \; \xi^{z-1}
\label{Differential}
\end{eqnarray}
Eq. (\ref{Differential}) should be supplemented by the boundary condition
\begin{eqnarray}
 \xi (x= - k_BT/f) = \xi_{\rm init} = \dfrac{k_BT}{f}
\label{Initial}
\end{eqnarray}
which represents the initial blob in the 'trumpet' (cf. Fig. \ref{Transient}).
Then, the solution of Eq. (\ref{Differential}) takes on the form
\begin{eqnarray}
 \xi(x) =  a \left[\dfrac{\zeta_0 \: v(t)}{k_BT} \left( x+\dfrac {k_BT}
{f}\right) + \left(\dfrac{a f}{k_BT}\right)^{z - 2} \right]^{1/(2-z)}
\label{Solution_Blob}
\end{eqnarray}
where in the case of Rouse dynamics $z-2= 1/\nu$.

One can fix the blob size by the requirement that the force at the free
boundary, i.e., at $x = - X(t)$, is equal to zero. Therefore,
\begin{eqnarray}
 f(x = - X(t)) = \dfrac{k_BT}{\xi(x = - X(t))} = 0\; ,
\label{Zero}
\end{eqnarray}
or, in view of Eq. (\ref{Solution_Blob}), this yields $(\zeta_0 \: v / k_BT) (-
X(t) + k_BT/f) + (a f/k_BT)^{z-2} = 0$. Solving the latter for $X(t)$,
\begin{eqnarray}
 X(t) = \dfrac{k_B T}{f} + \dfrac{k_B T}{\zeta_0 \: v(t)} \: \left(\dfrac{a
f}{k_BT}\right)^{z - 2} \; , \label{Border}
\end{eqnarray}
one may exclude $f$ in Eq. (\ref{Solution_Blob}) in
favor of $X(t)$. As a result we have
\begin{eqnarray}
 \xi(x) =  \dfrac{a}{ \left[\zeta_0 \: v(t) \: (x+X(t)) /k_B T
\right]^{1/(z-2)}} 
\label{Solution_Blob_1}
\end{eqnarray}
The largest blob of size $\xi_{X}(t) $ in the trumpet is placed at $x =
-X(t) + \xi_{X}(t)$. From Eq. (\ref{Solution_Blob_1}) one immediately obtains
\begin{eqnarray}
 \xi_{X}(t) = a \left[\dfrac{k_BT}{\zeta_0 \: a \; v(t)}\right]^{1/(z-1)}
\label{Blob_Max}
\end{eqnarray}

We are now in a position to derive the differential equation for the location of
the tension front $X(t)$. To this end we use the material balance equation
(mass conservation law). Let the tension be transmitted up to the $N(t)$-th
monomer whereas $M(t)$ denotes the number of translocated monomers (see Fig.
\ref{Transient}). Then the material balance equation reads
\begin{eqnarray}
 \int\limits_{-X(t)}^{0} \: \left[ \dfrac{\xi(x)}{a}\right]^{1/\nu}\: \dfrac{d
x}{\xi(x)} + M(t) = N(t) \label{Material_Balance}
\end{eqnarray}
where the first integral yields the number of monomers in the shaded portion
of the chain (see Fig.\ref{Transient}).

Denote the total number of monomers that has been subjected to tension during
the time interval $\{0, t\}$ by $N(t)$. At $t = 0$ all these $N(t)$ monomers
have been in equilibrium occupying a region of average size $X(t)$. Therefore
$N(t)$ and $X(t)$ are related by the Flory expression, i.e.,
\begin{eqnarray}
 X(t) = a N(t)^{\nu},
\label{Closure}
\end{eqnarray}
which will be used below as a closure relation.

The local density in the moving domain of blobs (shaded portion in Fig.
\ref{Transient}) is $\rho (x) \simeq g(x)/\xi(x)^3$ whereas the cross-section
area is locally $\Sigma (x) \simeq \xi(x)^2$. The smallest blob next to
the membrane orifice seeps through the pore into the {\it trans}-side of space
so that the number of translocated monomers can be calculated\cite{Sakaue_1} as
\begin{eqnarray}
 M(t) &=& \int\limits_{0}^{t} \left.\rho (x) \Sigma (x) \right|_{x=0}  v(t') \:
d t' = \int\limits_{0}^{t} \left[ \dfrac{\xi(x=0)}{a}\right]^{1/\nu}
\dfrac{1}{\xi(x=0)} \: v(t') \: d t'\nonumber\\ &=& \dfrac{1}{\tau_0} \:
\int\limits_{0}^{t}  \left[{\widetilde v}(t')\right]^{1 -
\frac{1-\nu}{\nu(z-2)}} \left[ {\widetilde X}(t')\right]^{- \frac{1 - \nu}{\nu(z
- 2)}}\: d t' \label{Trans_Monomers}
\end{eqnarray}
where we have used Eq. (\ref{Solution_Blob_1}) for the blob size and have
introduced the bead characteristic time $\tau_0 = a^2 \zeta_0/k_BT$ as well as
the dimensionless (tilded) variables \cite{Sakaue_2}: ${\widetilde v} (t) =
\zeta_0 a v(t)/k_BT$ and ${\widetilde X}(t) = X(t)/a$. Moreover, using  Eq.
(\ref{Border}) we can exclude ${\widetilde v}(t')$ in favor of ${\widetilde
X}(t)$ and the dimensionless force ${\widetilde f}_a = a f/k_BT$. As a result
\begin{eqnarray}
M(t)  = {\widetilde f}_a^{z - 1 -1/\nu}\: \dfrac{1}{\tau_0} \:
\int\limits_{0}^{t} \dfrac{d t'}{{\widetilde X}(t')\left[ 1 - 1/{\widetilde f}_a
{\widetilde X}(t')\right]^{1 - \frac{1-\nu}{\nu(z-2)}}} \label{M_VS_Time}
\end{eqnarray}

Now we use  Eqs. Eq. (\ref{Border}) and (\ref{Solution_Blob_1}) to determine the
first integral-term in Eq. (\ref{Material_Balance}) where we also take into
account Eq. (\ref{M_VS_Time}) and the closure Eq. (\ref{Closure}). As a result,
the material balance condition Eq. (\ref{Material_Balance}) takes on the form
\begin{eqnarray}
 {\widetilde X}(t)\left[ 1 - \dfrac{1}{{\widetilde f}_a {\widetilde
X}(t)}\right]^{\frac{1-\nu}{\nu(z-2)}} {\widetilde f}_a^{1 - \frac{1}{\nu}}  +
{\widetilde f}_a^{z - 1 - \frac{1}{\nu}} \: \dfrac{1}{\tau_0}
\int\limits_{0}^{t} \dfrac{d t'}{{\widetilde X}(t')\left[ 1 -
\frac{1}{{\widetilde f}_a {\widetilde X}(t') }\right]^{1 - \frac{1-\nu}{\nu
(z-2)}}} = \left[{\widetilde X}(t)\right]^{\frac{1}{\nu}}
\label{Balance_3}
\end{eqnarray}
For relatively long time intervals ${\widetilde f}_a {\widetilde X}(t) \gg
{\widetilde f}_a \simeq  1$ and Eq. (\ref{Balance_3}) takes on the form
\begin{eqnarray}
{\widetilde X}(t) \:{\widetilde f}_a^{1 - \frac{1}{\nu}}   + {\widetilde f}_a^{z
- 1 - \frac{1}{\nu}} \: \dfrac{1}{\tau_0} \int\limits_{0}^{t} \dfrac{d
t'}{{\widetilde X}(t')} = \left[{\widetilde X}(t)\right]^{\frac{1}{\nu}}
\label{Balance_4}
\end{eqnarray}

Differentiation of Eq. (\ref{Balance_4}) with respect to time yields a
differential equation for ${\widetilde X}(t)$
\begin{eqnarray}
 \tau_0 \left[1 - B_0 \left( {\widetilde f}_a {\widetilde X}(t) \right)^{1/\nu -
1}   \right] \: \dfrac{d {\widetilde X}(t)}{d t} = - \dfrac{{\widetilde f}_a^{z
- 2}}{{\widetilde X}(t)}
\label{Transmission_Diff}
\end{eqnarray}
where $B_0$ is a constant of the order of  unity. Eq.(\ref{Transmission_Diff})
should be supplemented by the initial condition
\begin{eqnarray}
 {\widetilde X}(t = 0) = {\widetilde \xi}_{\rm in} = \dfrac{1}{{\widetilde f}_a}
\end{eqnarray}
which simply indicates that the force transmission starts right after
the initial blob has formed as we have discussed in  Sec. IIA.  The solution of
Eq. (\ref{Transmission_Diff}) has the following form
\begin{eqnarray}
 t = t_0 + \tau_0 \:  B_0 \: {\widetilde f}_a^{1/\nu - z +1} \: {\widetilde
X}(t)^{1/\nu + 1} \lbrace  1 - C_0/\left[{\widetilde f}_a
{\widetilde X}(t)\right]^{1/\nu - 1}\rbrace \label{Solution_Time}
\end{eqnarray}
where $C_0 = 1/B_0$ and $t_0 = \tau_0 (1 - B_0)/{\widetilde f}_a^z$. The
characteristic time $\tau_1$ for the last monomer to attain its steady state
follows from ${\widetilde X}(\tau_1) = N^{\nu}$  which, due to  Eq.
(\ref{Solution_Time}), yields
\begin{eqnarray}
 \tau_1 = \tau_0 \: B_0 \: {\widetilde f}_a^{1/\nu - z +1} N^{1 + \nu}
\left[ 1 - C_0/{\widetilde f}_R^{1/\nu - 1}  \right]
\label{Tau_1}
\end{eqnarray}
where $N$ is the total chain length, ${\widetilde f}_R = R f/k_BT$ and $R = a
N^\nu$ stands for the unperturbed chain size.

\subsubsection{Stationary part of the translocation: suction into the pore}

The stationary regime sets in within a characteristic time $\tau_1$. After that
the non-translocated part of the chain, $N - M(\tau_1)$, is pulled as a whole
entity towards the pore (suction). The stationary velocity $v_s$ is defined from
Eq. (\ref{Border}) by putting ${\widetilde X}(\tau_1) = N^{\nu}$. As a result
for ${\widetilde f}_a {\widetilde X}(\tau_1) \gg 1$ we have
\begin{eqnarray}
 {\widetilde v}_s = \dfrac{1}{N^{\nu}} \: {\widetilde f}_a^{z - 2}
\label{Velocity_Stationar}
\end{eqnarray}
where the dimensionless velocity ${\widetilde v}_s = \zeta_0 \: a \: v_s/k_BT$.
The characteristic time for the stationary part of translocation is then
derived as
\begin{eqnarray}
 \tau_2 = \tau_0 \: \dfrac{{\widetilde X}(\tau_1) }{{\widetilde v}_s } = \tau_0
\: \dfrac{N^{2 \nu}}{{\widetilde f}_a^{z - 2}}
\label{Tau_2}
\end{eqnarray}

Eventually, the total translocation time $\langle \tau \rangle$ can be seen as
a sum of $\tau_1$ and $\tau_2$. Thus,  by taking into consideration Eq.
(\ref{Tau_1}) (at ${\widetilde f}_R \gg 1$) and Eq. (\ref{Tau_2}) we derive one
of our main results,
\begin{eqnarray}
 \langle \tau \rangle =  \tau_0 \: C_1 {\widetilde f}_a^{1/\nu - z +1} N^{1 +
\nu} + \tau_0 \: C_2 \: {\widetilde f}_a^{2 - z} \:  N^{2 \nu}
\label{Tau_Translocation}
\end{eqnarray}
where $C_1$ and $C_2$ are some numerical model dependent constants. The first
term in Eq. (\ref{Tau_Translocation}) dominates in case $C_1 {\widetilde
f}_a^{1/\nu - z +1} N^{1 + \nu} \gg C_2 \: {\widetilde f}_a^{2 - z} \:  N^{2
\nu}$, that is, for sufficiently large driving forces $f$ and moderate chain
lengths $N$, or, alternatively for very long chains even if the driving force
is rather weak:
\begin{eqnarray}
 {\widetilde f}_R \gg (C_2/C_1)^{\nu/(1 - \nu)}
\label{Condition}
\end{eqnarray}
where ${\widetilde f}_R = a N^{\nu} f/k_BT$. The combined schematic
representation of the scaling law, Eq. (\ref{Tau_Translocation}), as well as the
criterion given by Eq. (\ref{Condition}) is given in Table \ref{tab:Label}
and in Fig. \ref{Schematic}a,~b.

The central result given by Eq. (\ref{Tau_Translocation}) predicts a scaling
relation for the mean translocation time $\langle \tau \rangle \propto
N^{\alpha} f^{-\delta}$ with a force-dependent exponent $\alpha$ that grows from
$2\nu$ to $1+\nu$ as the parameter $a N^{\nu} f/k_BT$ increases. Numerical
results obtained by means of Langevin Dynamics (LD) simulation in two dimensions
(2D) clearly support this behavior: one observes $\alpha = 1.50$ for relatively
short chains ($N \leqslant 200$) and $\alpha = 1.69$ for longer chains
\cite{Luo_1}. In three dimensions (3D) there is no clear evidence of such
crossover. In the total interval of chain lengths which was studied in 3D the
exponent $\alpha = 1.41$ \cite{Luo_1} and $\alpha = 1.36$ \cite{Bhattacharya} ,
i.e. the exponent value falls in the range  between $2\nu$ and $1 + \nu$. In a
series of LD simulation studies by Lehtola et al.
\cite{Lehtola_1,Lehtola_2,Lehtola_3} it was found (within chain length interval
$N \leqslant 800$) that $\alpha$ systematically increases with $f$ up to roughly
$\alpha = 1+\nu$. This is in agreement with our theoretical and MD-results (see
below Sec. \ref{Numerics}) but is in clear contradiction to the results of K.
Luo et al. \cite{Luo_2} where the exponent $\alpha = 1+\nu \approx 1.59$ was
found for relatively short chains, $N \leqslant 200$ and for small driving
forces. Moreover, the exponent $\alpha$ becomes smaller, $\alpha \approx 1.37$
as the driving force grows. Experimentally, the scaling law $\langle \tau
\rangle \propto N^{2\nu}/f$, which follows from Eq. (\ref{Tau_Translocation})
for relatively weak forces ${\widetilde f}_R$ and $z = 3$ (Zimm dynamics), has
been found in ref. \cite{Dekker_1,Dekker_2} in the case of synthetic pore.
Recently the scaling given  by the first term in Eq. (\ref{Tau_Translocation})
has been obtained in ref. \cite{Grosberg} using the so-called ``iso-flux
trumpet'' model (which was also inspired by the Sacaue's paper). 

Eventually, we should like to point out that the first term of Eq.
(\ref{Tau_Translocation}) (which corresponds to the characteristic time of
tensile force transmission) {\em differs} from the corresponding term given by
Eq. (10) in ref.\cite{Sakaue_2}. This difference arises even if one assumes,
following Sakaue's idea, that the entire trumpet is characterized by a single
time dependent velocity (``iso-velocity trumpet''!). The reason for this
discrepancy is discussed in the Appendix \ref{App}.

\subsubsection{Time evolution of the translocated portion $M(t)$}

The number of translocated monomers $M(t)$, given by Eq. (\ref{M_VS_Time} ),
grows with elapsed time. In the case when the second term in Eq.
(\ref{Solution_Time}) dominates, one gets
\begin{eqnarray}
 t \simeq \tau_0 \:  B_0 \: {\widetilde f}_a^{1/\nu - z +1} \: {\widetilde
X}(t)^{1/\nu + 1}
\end{eqnarray}
Combining this equation with Eq. (\ref{M_VS_Time}) (where again we assume
${\widetilde f}_a {\widetilde X}(t) \gg 1$), one arrives at
\begin{eqnarray}
 M(t) = ({\widetilde f}_a)^{\chi} \; \int\limits_{0}^{t/\tau_0} \;
({\widetilde t})^{-\frac{\nu}{1+\nu}} \: d \: {\widetilde t} =  C_0
\: ({\widetilde f}_a)^{\chi} \: ({\widetilde t})^{1/(1+\nu)}
\label{Translocation_Law}
\end{eqnarray}
where the exponent $\chi = (z \nu -1 - \nu)/\nu (1+\nu)$. In the case of Rouse
dynamics $z\nu = 2\nu +1$ and $\chi = (1+\nu)^{-1}$. The resulting translocation
relationship can be written as
\begin{eqnarray}
M(t)  = c_0 \left( {\widetilde f}_a \:  {\widetilde t}\right)^{1/(1 + \nu)}
\label{Translocation_Law_Rouse}
\end{eqnarray}
As far as mean translocation time $\langle \tau \rangle$ should satisfy the
requirement $M(\langle \tau \rangle) = N$, one recovers the result $\langle
\tau \rangle \simeq \tau_0 N^{1+\nu}/f$ (see Eq. (\ref{Tau_Translocation})).

In the case of small driving force, ${\widetilde f}_a N^{\nu} \simeq 1$ and
${\widetilde f}_a N^{\nu} \ll (C_2/C_1)^{\nu/(1-\nu)}$, the translocation
process is mainly determined by the stationary suction into the pore whereby the
translocation time  $\langle \tau \rangle \simeq C_2 \tau_0 N^{2 \nu}/
{\widetilde f}_a^{z-2}$. On the other hand, the number of translocated monomers
according to Eq. (\ref{Trans_Monomers}) reads
\begin{eqnarray}
M(t)  = {\widetilde v}_s \: \int\limits_{0}^{t} \left[
\dfrac{\xi(x=0)}{a}\right]^{1/\nu} \dfrac{1}{\xi(x=0)}  \: d t = [ {\widetilde
v}_s ]^{1 - \frac{1 - \nu}{\nu (z-2)}} \dfrac{1}{\tau_0} \: \int\limits_{0}^{t}
\; \dfrac{d t'}{[{\widetilde X} (t')]^{\frac{1 - \nu}{\nu (z-2)}} }
\label{Trans_Monomers_Stat}
\end{eqnarray}
where we have used the expression for the blob size Eq. (\ref{Solution_Blob_1})
with the  stationary velocity ${\widetilde v}_s$. In the stationary regime the
size of the moving domain has the form ${\widetilde X} (t) = {\widetilde X}
(\tau_1) - {\widetilde v}_s t/\tau_0 \simeq N^{\nu} - {\widetilde v}_s
t/\tau_0$. By making use of this in Eq. (\ref{Trans_Monomers_Stat}) and after
integration, we arrive at the result
\begin{eqnarray}
 M(t) = \dfrac{N^{\nu}}{{\widetilde f}_a^{(1 - \nu)/\nu}} \: \left\lbrace 1 -
\left[1 - \frac{t}{\tau_2}  \right]^{1 - \frac{1 - \nu}{\nu (z - 2)}}
\right\rbrace \label{Trans_Monomers_Stat_1}
\end{eqnarray}

Consider first the large time limit $t \simeq \tau_2$. In this case $M(\tau_2)
\simeq N$ and the Eq. (\ref{Trans_Monomers_Stat_1}) leads us back to the
condition ${\widetilde f}_a N^{\nu} \simeq 1$. At $t \ll \tau_2$ the expansion
in Eq. (\ref{Trans_Monomers_Stat_1}) suggests that
\begin{eqnarray} M(t) \simeq
c_1 \: \dfrac{{\widetilde f}_a^{z - 1 - 1/\nu}}{N^{\nu}} \: {\widetilde t}
\end{eqnarray}
where, as before, ${\widetilde t} = t/\tau_0$. In the case of Rouse dynamics we
finally obtain
\begin{eqnarray}
 M(t) \simeq  \dfrac{c_1}{N^{\nu}}  \: {\widetilde f}_a {\widetilde t}  
\label{Trans_Monomers_Stat_2}
\end{eqnarray}

As can be seen from Eq. (\ref{Trans_Monomers_Stat_2}), the formal extrapolation
the linear time dependence law up to the final time $t_{f}$ (where  $M (t_f)
\simeq N$) gives $t_f \propto N^{1+\nu}/{\widetilde f}_a \simeq  \tau_1 \ll
\tau_2 $. As a consequence, a sub-linear slowing down behavior should be seen at
a later time interval. The two regimes mentioned above (i.e., for strong and
weak driving forces) are illustrated in Fig. \ref{Schematic}c.

\subsection{``Stem'' and ``stem-trumpet'' scenario}
\label{Stem_Section}

\subsubsection{Stem}
The above consideration is valid in the range of  forces given by Eq.
(\ref{Forces_Range}). For stronger forces which are within the range
\begin{eqnarray}
 1 \leqslant {\widetilde f}_a \ll N^{\nu}
\label{Stem_Flower_Force_Interval}
\end{eqnarray}
the tensile force transmission follows the so called {\it stem-trumpet} scenario
\cite{Brochard_1,Brochard_2}, as indicated in Fig. \ref{Transient}b, whereby the
part of the polymer chain that is close to the membrane attains a completely
stretched conformation.

Consider first the limit of  very strong force ${\widetilde f}_a > N^{\nu}$
where the configuration is depicted by the {\it stem} picture shown in Fig.
\ref{Transient}c. In this case the force balance reads
\begin{eqnarray}
 {\widetilde v} (t) {\widetilde X}(t) = {\widetilde f}_a
\label{Force_Balance}
\end{eqnarray}
The material balance (cf. Eq. (\ref{Material_Balance})) yields
\begin{eqnarray}
 {\widetilde X}(t) + M(t) = N(t)
\label{Material_Balance_1}
\end{eqnarray}

The number of translocated segments (cf. Eq.(\ref{Trans_Monomers})) is
\begin{eqnarray}
 M(t) = \dfrac{1}{a} \int\limits_{0}^{t} v(t') d t' = \dfrac{1}{\tau_0}
\int\limits_{0}^{t} {\widetilde v}(t') d t'
\label{Trans_Monomers_1}
\end{eqnarray}
With the help of Eqs. (\ref{Force_Balance}), (\ref{Material_Balance_1}) and
(\ref{Trans_Monomers_1}) as well as using the closure Eq. (\ref{Closure}) one
finds
\begin{eqnarray}
 {\widetilde X} (t) + \dfrac{{\widetilde f}_a}{\tau_0} \; \int\limits_{0}^{t}
\dfrac{ d t'}{{\widetilde X} (t')} = \left[{\widetilde X} (t) \right]^{1/\nu}
\end{eqnarray}
In differential form this equation reads
\begin{eqnarray}
 \tau_0 \left\lbrace C_0 \left[  {\widetilde X} (t)\right]^{1/\nu} - C_1
{\widetilde X} (t) \right\rbrace \: \dfrac{d {\widetilde X} }{d t} = {\widetilde
f}_a \label{Differential_X}
\end{eqnarray}
with an initial condition ${\widetilde X} (t=0) = 1$ and $C_0$, and $C_1$ being
some constants. The corresponding solution of Eq. (\ref{Differential_X}) is
given by
\begin{eqnarray}
\dfrac{{\widetilde f}_a}{\tau_0} t - {\widetilde C}_1  + {\widetilde C}_0 =
{\widetilde C}_0 \left[  {\widetilde X} (t)\right]^{1/\nu + 1} \left \{ 1 -
\dfrac{{\widetilde C}_1}{ {{\widetilde C}_0 \left[  {\widetilde X}
(t)\right]^{1/\nu - 1}}} \right\} \label{Solution_Time_1}
\end{eqnarray}
Again the characteristic time $\tau_1$ is defined as ${\widetilde X} (\tau_1) =
{\widetilde R} = N^{\nu}$. Then, for the reasonably long chains $N^{1-\nu} \gg
{\widetilde C}_1/{\widetilde C}_0$ the first terms in Eq.(\ref{Solution_Time_1})
prevails. As a result, the characteristic time $\tau_1$ is obtained as
\begin{eqnarray}
 \tau_1 = \tau_0 \: \dfrac{N^{1+\nu}}{{\widetilde f}_a}
\label{Tau_1_Stem_Coil}
\end{eqnarray}

As before, the second (stationary) stage of the translocation is related to the
pulling force by considering this part of the chain as an entity which moves
with stationary velocity ${\widetilde v}_s = {\widetilde f}_a/N^{\nu}$. The
corresponding characteristic time $\tau_2$ is given by
 \begin{eqnarray}
  \tau_2 = \tau_0 \: \dfrac{{\widetilde R}}{{\widetilde v}_s} = \tau_0 \:
\dfrac{N^{2 \nu}}{{\widetilde f}_a}
 \end{eqnarray}
Therefore, the total translocation time $\langle \tau \rangle$ is a combination
of $\tau_1$ and $\tau_2$, i.e.,
\begin{eqnarray}
 \langle \tau \rangle = C_1 \: \tau_0 \dfrac{N^{1+\nu}}{{\widetilde f}_a} +
C_2 \: \tau_0 \: \dfrac{N^{2 \nu}}{{\widetilde f}_a}
\label{Transloc_Time_Stem_Coil}
\end{eqnarray}
The first term in Eq. (\ref{Transloc_Time_Stem_Coil}) dominates at
\begin{eqnarray}
 N^{1 - \nu} \gg C_2/C_1
\label{Long_Chain}
\end{eqnarray}
i.e., for a reasonably long chains. One should note that in this limit we
recover the scaling law of the trumpet regime, provided Eq. (\ref{Long_Chain})
holds and ${\widetilde X}(t) = ({\widetilde f}_a {\widetilde t})^{\nu/(1 +
\nu)}$. The velocity ${\widetilde v} (t) = {\widetilde f}_a/{\widetilde X}(t)=
{\widetilde f}_a ({\widetilde f}_a {\widetilde t})^{- \nu/(1+\nu)}$, and the
number of translocated monomers is obtained as
\begin{eqnarray}
 M(t) = \dfrac{1}{a} \int\limits_{0}^{t} {\widetilde v}(t') d t' \simeq
B_0 \: \left( {\widetilde f}_a \; {\widetilde t}\right)^{1/(1 + \nu)}
\label{Translocation_Law_Stem_Coil}
\end{eqnarray}
whereby one goes back to Eq. (\ref{Translocation_Law_Rouse}).

\subsubsection{The Stem-trumpet regime}

In the interval of force strengths, Eq. (\ref{Stem_Flower_Force_Interval}), the
chain deformation starts with the formation of a stem whereby the velocity
changes as
\begin{eqnarray}
 {\widetilde v}(t) \simeq \left({\widetilde f}_a \right)^{1/(1+\nu)}\;
\left(\dfrac{\tau_0}{t} \right)^{\nu/(1 + \nu)}
\end{eqnarray}
Thus, the velocity decreases with time and at some moment $t \simeq \tau^{\#}$
the drag force at the stem-trumpet junction becomes comparable to $k_BT/a$,
i.e., $\zeta_0 v(\tau^{\#}) \simeq k_BT/a$. Therefore,
\begin{eqnarray}
 \tau^{\#} \simeq \tau_0 \; ({\widetilde f}_a)^{1/\nu}
\end{eqnarray}

At $t > \tau^{\#}$ the stem-trumpet picture dominates. For the blob size of the
flower (trumpet) part the same differential equation Eq. (\ref{Differential})
holds. However, the boundary conditions (BC) are different. Namely, in this case
$\xi (x = - S(t)) = a$ and the solution of Eq.(\ref{Differential}) yields
$\xi(x) = a \left\lbrace 1 + \zeta_0 v(t)[x+S(t)]/T\right\rbrace ^{1/(2-z)}$.
Therefore, the solution is given in the total $x$-interval by
 \begin{eqnarray}
  \xi(x) = \begin{cases}
           a \left\lbrace 1 + \zeta_0 v(t) [x + S(t)]/k_BT
\right\rbrace^{1/(2-z)} \; &, \;\mbox{for} \quad - X(t) \leq x \leq -
S(t)\\
a \quad &,\;\mbox{for} \quad - S(t) \leq x \leq 0
          \end{cases}
\label{Blob_Size_Stem_Flower}
\end{eqnarray}
For $x = - X(t)$ the force vanishes, i.e., $f(x = - X(t)) = T/\xi(x =
-X(t)) = 0$, and one has
\begin{eqnarray}
 X(t) = S(t) + \dfrac{k_BT}{\zeta_0 v(t)}
\label{X_S_v}
\end{eqnarray}

The material balance may be written as
\begin{eqnarray}
 \int\limits_{-X(t)}^{-S(t)} \: \left[\dfrac{\xi(x)}{a}\right]^{1/\nu} \dfrac{d
x}{\xi(x)} + \dfrac{1}{a} \; S(t) + M(t) = N(t)
\label{Material_Balance_Stem_Flower}
\end{eqnarray}
By making use of Eqs. (\ref{Closure}), (\ref{Blob_Size_Stem_Flower}) and
(\ref{X_S_v}) in Eq. (\ref{Material_Balance_Stem_Flower}), one arrives at
\begin{eqnarray}
 {\widetilde X}(t) + \dfrac{1}{a} \int\limits_{0}^{t} v(t') d t' = \left[
{\widetilde X}(t)\right]^{1/\nu} \label{Stem_Flower_Balance}
\end{eqnarray}
In order to exclude the velocity from Eq. (\ref{Stem_Flower_Balance}) we should
note that the tensile force at $x = -S(t)$ (junction point between ``stem'' and
``trumpet'') is given by
\begin{eqnarray}
 f(x= - S(t)) = \dfrac{k_BT}{a}
\end{eqnarray}
Then the force balance for the stem can be written as
\begin{eqnarray}
 \zeta_0 v(t) \left[  \dfrac{S(t)}{a}\right] = f - f(x= - S(t))
= f - \dfrac{k_BT}{a}\; ,
\end{eqnarray}
and consequently,
\begin{eqnarray}
 {\widetilde S}(t) = \dfrac{1}{{\widetilde v}(t)} \: ({\widetilde f}_a - 1)
\end{eqnarray}
This equation together with Eq. (\ref{X_S_v}) yields
\begin{eqnarray}
 {\widetilde X}(t) = \dfrac{1}{{\widetilde v}(t)} \:{\widetilde f}_a
\label{X_v_f}
\end{eqnarray}
and one retrieves the corresponding equation Eq.(\ref{Force_Balance}) for the
stem case. Now, using  Eq. (\ref{X_v_f}) in Eq. (\ref{Stem_Flower_Balance})
one arrives at the differential equation for ${\widetilde X}(t)$
\begin{eqnarray}
 \tau_0 \left\lbrace B_0 \left[ {\widetilde X}(t)\right]^{1/\nu} - B_1
{\widetilde X}(t) \right\rbrace \: \dfrac{d {\widetilde X}(t)}{d t} =
{\widetilde f}_a
\end{eqnarray}
which is again equivalent to Eq. (\ref{Differential_X}) for the ``stem'' case.
As a consequence, the expressions for the translocation time and $M(t)$
are given by Eqs. (\ref{Transloc_Time_Stem_Coil}) and
(\ref{Translocation_Law_Stem_Coil}) respectively.

The main conclusion of this Sec. \ref{Stem_Section} lies in the fact that for
the moderate and strong forces the translocation exponent  changes from $\alpha
= 2\nu$ for moderately long chains to $\alpha = 1 + \nu$ for very large values
of $N$. This is in good correspondence with the results of the Monte Carlo
investigation \cite{Slater}.

\section{Numerical verification of theoretical predictions}
\label{Numerics}

\subsection{Model}

In order to verify the predictions of Section \ref{sec:theory} we performed
numerical simulations. The model assumes Langevin dynamics of a polymer chain
which consists of $N$ beads that thread through an octagonal pore through a
closely-packed wall (membrane). The interaction between the monomers of the
chain is modeled by a Finitely Extensible Nonlinear Elastic (FENE) springs
corresponding to a pair potential
\begin{align}
 U_{FENE}(r_{ij})&=-\frac{k r_{ij}^2}{2} \ln\left(1-\frac{r_{ij}^2}{R_0^2}\right),
\end{align}
where $r_{ij}$ is the bond length between two beads and $R_0=1.5$ is the maximal
bond length. All beads  experience excluded volume interactions which are
modeled by the repulsive part of the shifted Lennard-Jones potential,
\begin{figure}[ht]
\includegraphics[scale=0.47]{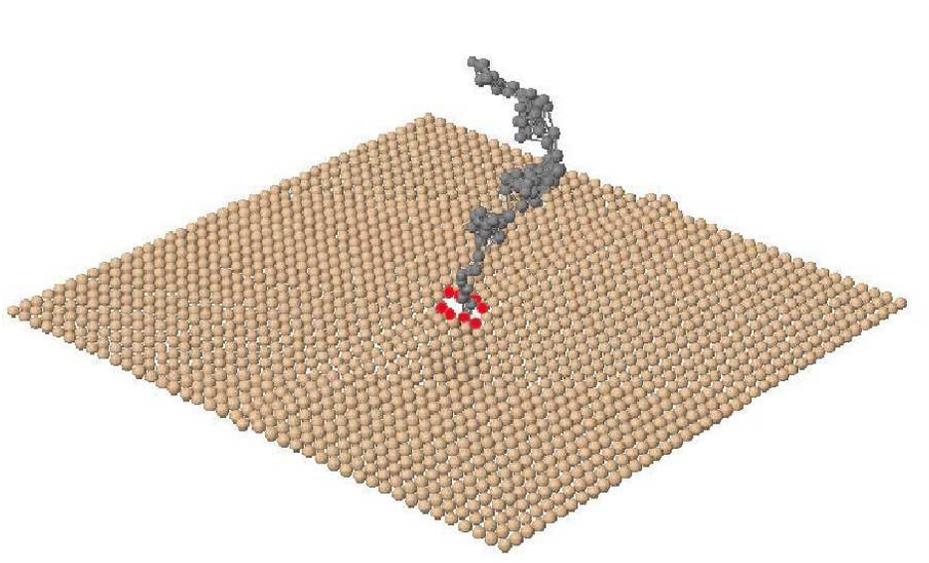}
\caption{(Color online) A snapshot configuration during translocation of the
polymer chain. The nanopore is created by removing 8 beads, so that an octagonal
hole results.}
\label{fig:3d}
\end{figure}
also known as the Weeks-Chandler-Anderson (WCA) potential. This potential
$U_{WCA}$ is defined by
\begin{align}
 U_{WCA}(r_{ij})=4\epsilon\left[\left(\frac{\sigma}{r_{ij}}\right)^{12}
-\left(\frac{\sigma}{r_{ij}}\right)^{6} + \frac{1}{4}\right] \Theta(r_c - r),
\end{align}
where $\Theta(x)$ is the Heavyside-function, i.e., we use a cut-off
$r_c=2^{-1/6}\sigma$, implying $U_{WCA}=0$ for $r_{ij}>r_c$. The monomers
residing inside the pore experience a constant external force $f$ in the
direction perpendicular to the membrane, which we designate by $x$. The external
force can be implemented by adding a linear potential $U_{ext}$, whose value is
$0$ outside the pore and  $f x$, if $x$ is inside the pore region. Thus, $f$
pulls the chain towards the region of positive $x$ which we refer to as the {\em
trans}-side. The equation of motion for the beads of the chain reads
\begin{align}
 m\frac{d^2\bf{r}_i}{dt^2}=-\nabla \left(U_{FENE}+U_{LJ}+U_{ext}\right) - \gamma
\frac {d {\bf r}_i} {dt} + {\cal R}_i(t), \end{align} where ${\cal R}_i(t)$
stands for a Brownian random force whose moments obey $\langle {\cal
R}_i(t)\rangle=0$ and $\langle {\cal R}_i^{\alpha} (t_1) {\cal
R}_j^{\beta}(t_2)\rangle = 2 k_B T \gamma
\delta(t_1-t_2)\delta_{\alpha,\beta}\delta_{i,j}$ The parameter values were set
to  $\epsilon=1.2$, $\sigma=1.0$, $k=60.0$, $\gamma=0.73$  and were kept fixed
during the simulations. The temperature had a constant value given by
$T=1.2\epsilon/k_B$. These parameters correspond to the model used in the
ref.\cite{Lehtola_3}.

The membrane is modeled by a plane of beads whose positions are kept fixed.
Eight neighboring monomers are removed to obtain an octagonal pore. The
interaction between the beads of the chain and the plane is mediated through the
repulsive WCA potential. All simulations were performed starting from a
configuration in which initially all beads except the first one  were on the
{\em cis}-side. The chain is fully translocated when all beads have made their
way to the {\em trans}-side. To prevent the chain from retracting to the {\em
cis}-side, reflecting boundary conditions were imposed on the first monomer on
the chain by making its size larger than the pore diameter. A typical
configuration for a polymer with $N=100$ in $3D$ is shown in Fig.~\ref{fig:3d}.

\subsection{Results}

\begin{figure}[ht]
%\subfigure[]{
\vspace{1.0cm}
\includegraphics[scale=0.45,angle=270]{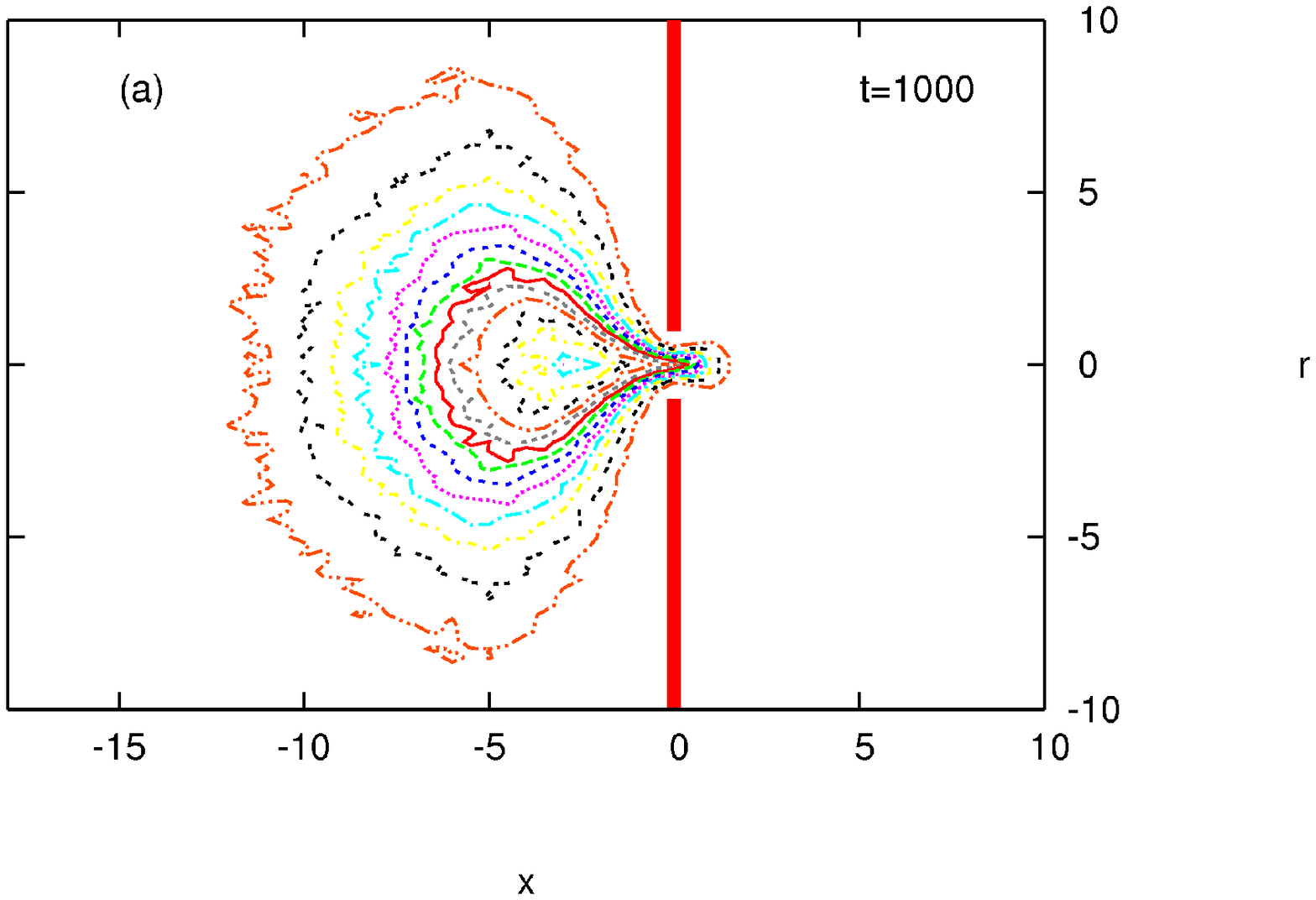}
%\caption{The densityplot corresponding to the initial configuration. The
%chaincomprises 100 beads and the driving force $F=2$.  The membrane has a hole
%at $z=-\frac{1}{2}$. All %beads are initially on the cis-side.}
%\label{fig:dens1}}
%\subfigure[]{
\hspace{1.0cm}
\includegraphics[scale=0.45,angle=270]{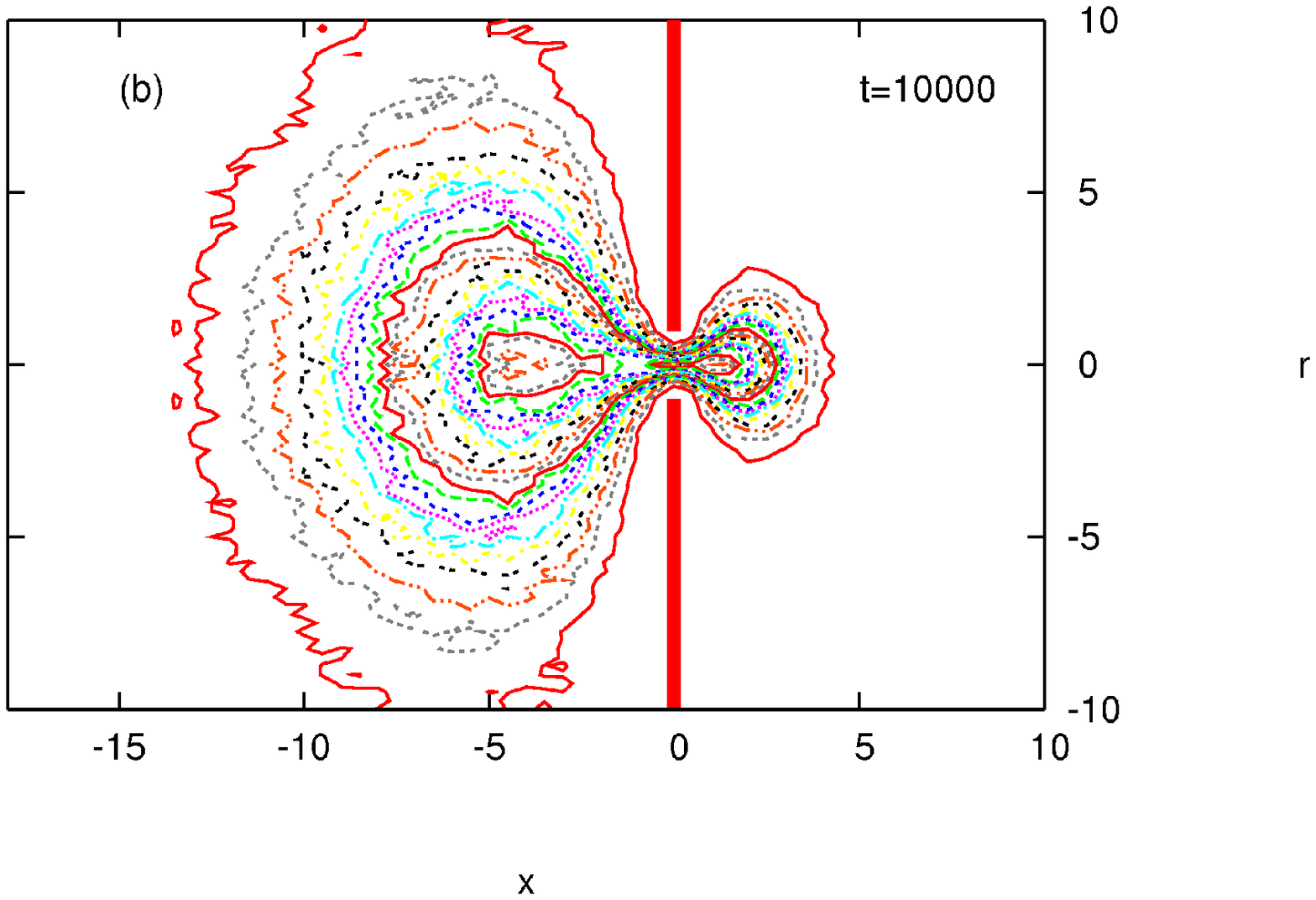}
%\caption{The densityplot corresponding to a configuration after full
%translocation was reached. The membrane has a hole at $z=-\frac{1}{2}$. All
%beads are now at the trans-side. Crowding can be seen.}
%\label{fig:dens2}
%}
%\subfigure[]{
\vspace{1.0cm}
\includegraphics[scale=0.45,angle=270]{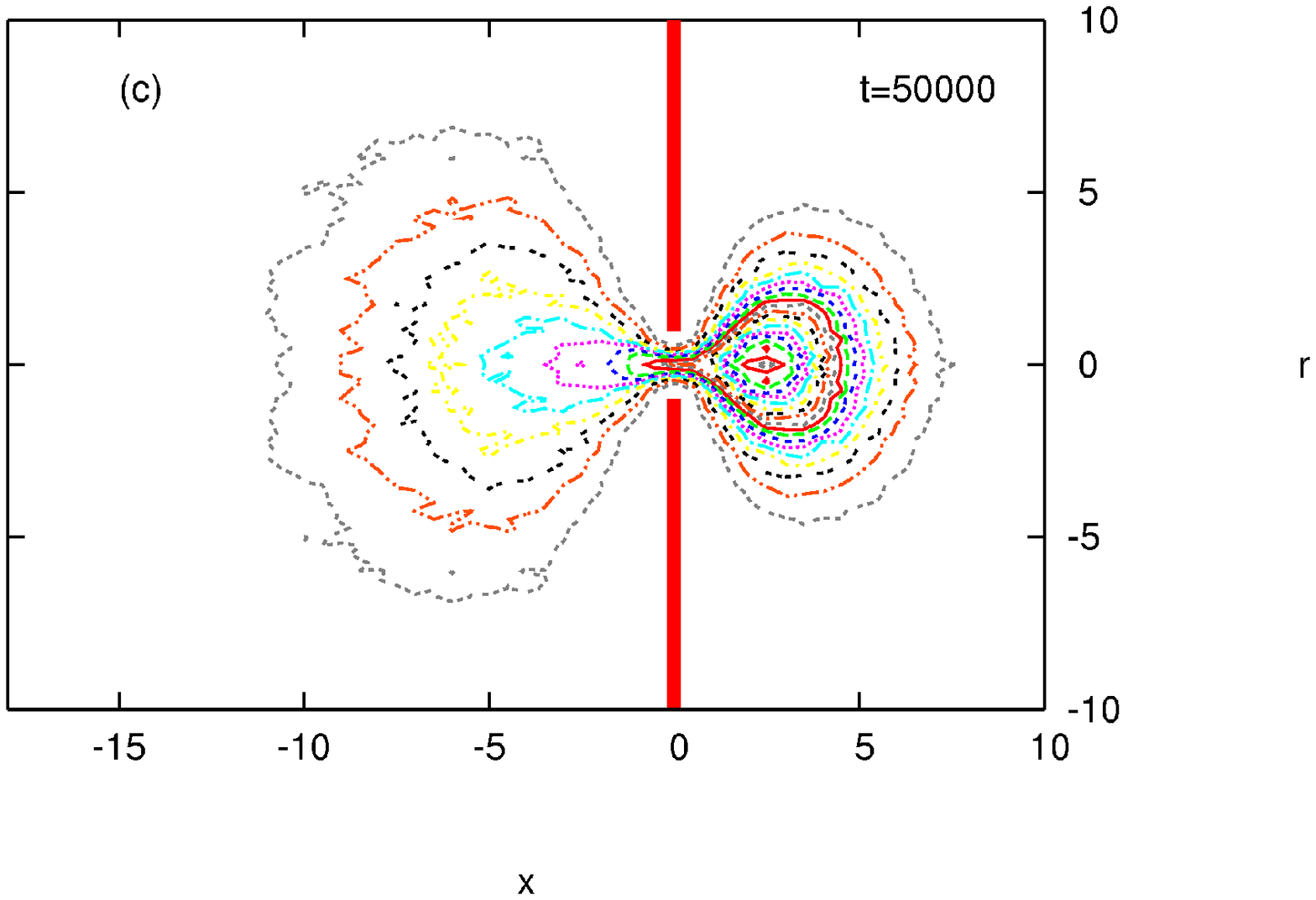}
%\caption{The densityplot corresponding to a configuration after full
%translocation was reached. The membrane has a hole at $z=-\frac{1}{2}$. All
%beads are now at the trans-side. Crowding can be seen.}
%\label{fig:dens3}
%}
%\subfigure[]{
\hspace{1.0cm}
\includegraphics[scale=0.45,angle=270]{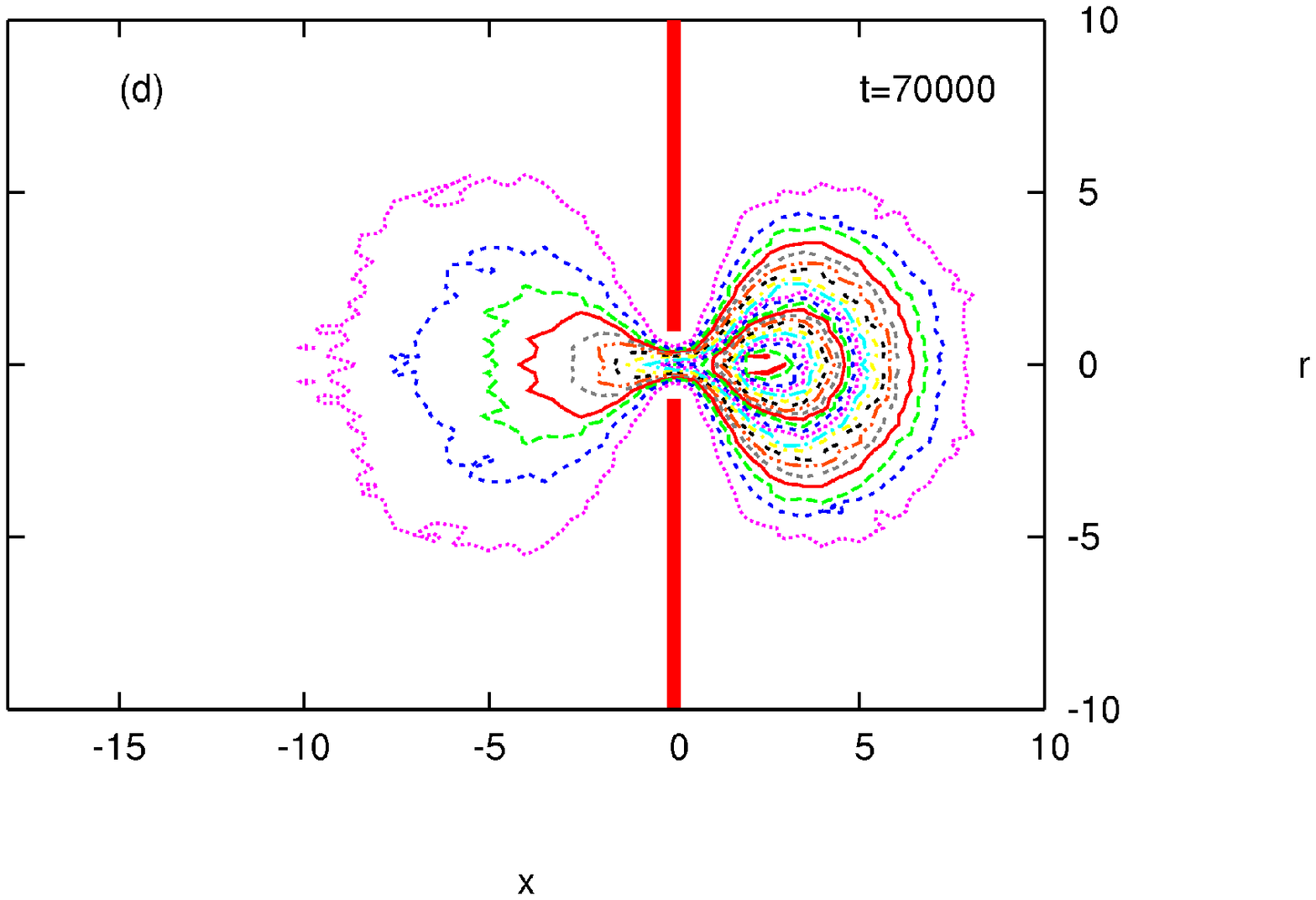}
%\caption{The densityplot corresponding to a configuration halfway the
%translocation process. The membrane has a hole at $z=-\frac{1}{2}$. All %beads
%are still mainly at the cis-side, but some have translocated to the trans-side.
%The density seems to reflect a %trumpet-like shape as discussed in Section.}
%\label{fig:dens4}
%}
%%\subfigure[]{
%%\includegraphics[scale=0.34]{densityplotN100F299.eps}
%\caption{The densityplot corresponding to a configuration after full
%translocation was reached. The membrane has a hole at $z=-\frac{1}{2}$. All
%beads are now at the trans-side. Crowding can be seen.}
%%\label{fig:dens5}
%%}
%%\subfigure[]{
%%\includegraphics[scale=0.34]{densityplotN100F2fin.eps}
%\caption{The densityplot corresponding to a configuration after full
%translocation was reached. The membrane has a hole at $z=-\frac{1}{2}$. All
%beads are now at the trans-side. Crowding can be seen.}
%%\label{fig:dens6}
%%}
\vspace{0.0cm}
\caption{(Color online) Contour plots of density  corresponding to different
time moments for the force $f = 2$ and the chain length $N = 100$. In (a) the
density profile is plotted at $t=1000$, (b)  is for $t=10^4$, (c) for $t=5\times
10^4$  and (d) for $t=7\times 10^4$. }\label{fig:dens1}
\end{figure}

To check the validity of the blob picture, derived in Sec.~\ref{sec:theory}, we
start by preparing contour  plots of the monomer density. We assume that the
chain propagates in the positive $x$-direction, that is, perpendicular to the
membrane; the $z$-direction is parallel to the membrane and equivalent to the
$y$. The contour  plots of the density distribution Figs.~\ref{fig:dens1}a -
\ref{fig:dens1}d describe the change in the average polymer  conformation at
different stages of the translocation process.

\begin{figure}[ht]
\includegraphics[scale=0.35]{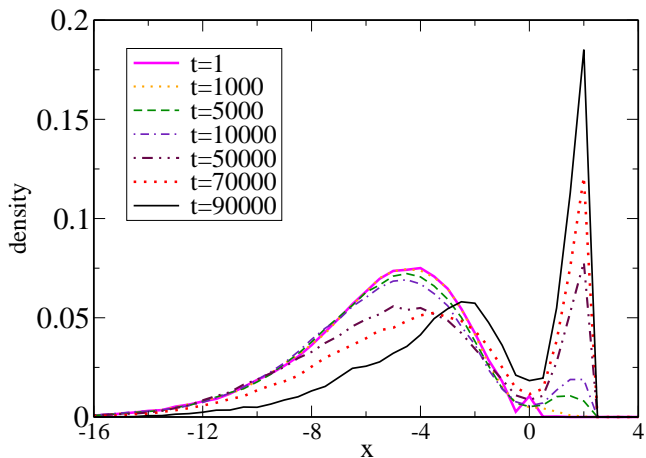}
\vspace*{1.5cm}
\caption{(Color online) Density profiles (normalized to unity ) as function
of $x$-coordinate. In the time interval $0 < t \le 70000$ the front moves toward
the left, i.e., the moving domain grows at the expense of the rear part of
the chain. At $t > 70000$ the second stage of the translocation, ``suction in
the pore'', sets in. On the  trans-side (i.e. $x>0$) the ``crowding'' effect can
be seen.}
\label{DensityVSz}
\end{figure}

In order to gain more insight  into the translocation process we have also
plotted similar data in a different form. Figure~\ref{DensityVSz} demonsrates
the density profiles (normalized to unity) in $x$-direction for different time
moments. This representation clearly shows that the tensile force transmission
(or front propagation) in $x$-direction (as it was described in Sec. IIC) is
really visible. Namely, the moving domain grows at the expense of the rear part
of the chain which is still at rest. After approximate time $t = 70000$, the
non-translocated part of the chain moves as a whole (cf. Sec. IIC1, where this
second stage of translocation was qualified as the ``suction into the pore''). 

Figure~\ref{DensityVSz} also clearly shows the presence of ``crowding'' effect,
i.e., a fairly strong compression of the coil on the {\em trans}-side in
$x$-direction (i.e. at $x > 0$). This effect results from the fact that the
characteristic time $\tau$ of a forced translocation (which ranges, as we have
shown in Sec. \ref{TR}, between $\tau_1 \sim N^{1+\nu}$ and $\tau_2 \sim
N^{2\nu}$), is by all means much smaller than the characteristic Rouse time
$\tau_R \sim N^{2\nu+1}$, that is, the threaded beads on the {\em trans}-side
fail to disperse and form an equilibrated coil conformation in an interval $\tau
< \tau_R$. This crowding effect has been discussed recently in more detail
\cite{Aniket} and it was shown that immediately after translocation an effective
Flory exponent $\nu_{eff} = 0.45$ is observed, i.e., $\nu_{eff}$  is smaller
than $\nu = 0.59$ for a chain in equilibrium.

\subsubsection{Scaling of the translocation time}

In Sec. \ref{TR} we have shown that the time for a driven translocation scales
as $\langle \tau \rangle \sim N^{\alpha} f^{-\delta}$. For large forces (the
corresponding condition is given by  Eq. (\ref{Condition})) the translocation
exponent $\alpha = 1+\nu \approx 1.59$ and $\delta = 1$ (for the so-called
free-draining or Rouse dynamics). Thus one goes back to the scaling
relationship, predicted first by Kantor and Kardar \cite{Kantor}. Very recently
\cite{Grosberg} the same scaling has been obtained on a basis  of the so-called
"iso-flux trumpet`` consideration. In fact, in this case the translocation time
is determined by the propagation of tensile force with characteristic time
$\tau_1$ given by Eq. (\ref{Tau_1}). At relatively small forces, in contrast,
the exponents are predicted to be $\alpha = 2\nu \approx 1.18$, and  $\delta =
1/\nu \approx 1.7$ (``trumpet'' regime), or $\delta = 1$ (``stem-trumpet''
regime). At small forces the process is dominated by stationary suction of the
rest of the chain after the force transmission stage.

\begin{figure}[ht]
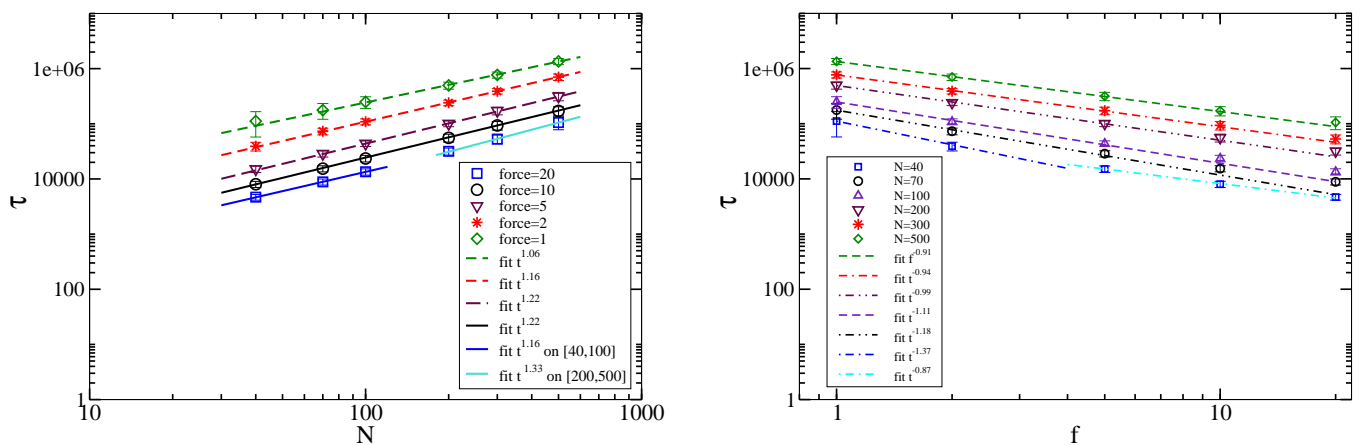

\vspace{0.9cm}
 \includegraphics[scale=0.34]{fig6a.eps}
\hspace{0.5cm}
 \includegraphics[scale=0.34]{fig6b.eps}
\caption{(Color online) (a) The $\tau$ vs $N$ for different forces $f$. The
figure shows that $\tau \sim N^{\alpha}$, where $\alpha$ depends on the force.
For small forces ($f=1,\; 2$), the value of $\alpha$ is about $1.1$, for larger
forces ($f=10,20$), a crossover appears at $N^{*}=N \approx 100$. For chain
lengths $N < N^{*}$ one has $\alpha\approx 1.16$ which is close to $2\nu$,
whereas for  $N > N^{*}$, $\alpha=1.33$. (b) $\tau$ vs $f$ for chain lengths
$N=40,\; 100,\; 200,\; 300,\; 500$. One finds $\tau \propto 1/f^\delta$ with
$\delta \approx 1$. For short chains lengths the exponent $\delta$ is somewhat
larger than $1$.} 
\label{fig:fig6}
\end{figure}

The results for the $\langle \tau \rangle$ vs. $N$ scaling relationship at
different driving forces, derived from our MD-simulation, are shown in Fig.
\ref{fig:fig6}a. We performed numerical simulations for chains with lengths
$N=40,70,100,200,300,500$ and forces $1,2,5,10,20$. For every set of parameters
at least $1000$ runs were performed, and most averages were obtained from $3000$
separate runs. The $N=500$ runs took a long time, mostly because relaxation to
equilibrium (as a starting condition) was very time consuming. The $f=1,2$
results for the $N=500$ chain are therefore typically averaged over $1000$ runs.
It can be seen from Fig.~\ref{fig:fig6}a that the theoretically predicted
tendency is correct: the translocation exponent $\alpha$ grows with increasing
force. Nevertheless, quantitatively  the MD-values are systematically smaller
than the  theoretically expected ones: $\alpha \approx 1.33$ for the large force
and $\alpha \approx 1.06$ for the small force. This behavior closely corresponds
to the results found by the Langevin MD method
\cite{Lehtola_1,Lehtola_2,Lehtola_3}.

Fig.~\ref{fig:fig6}b shows how the translocation time changes with force. For
the long chains the exponent $\delta$ is very close to unity in a good agreement
with the theoretical prediction. For a shorter chain $\delta \approx 1.37$
which is much smaller than the possible theoretical prediction within the
trumpet scenario: $\delta \approx 1.7$.

%\begin{figure}[ht]

%\caption{The translocation time as a function of the driving force $f$}
%\label{fig:fig7}
%\end{figure}

\subsubsection{Evolution of the translocation coordinate}

\begin{figure}[ht]
 \includegraphics[width =9.0cm]{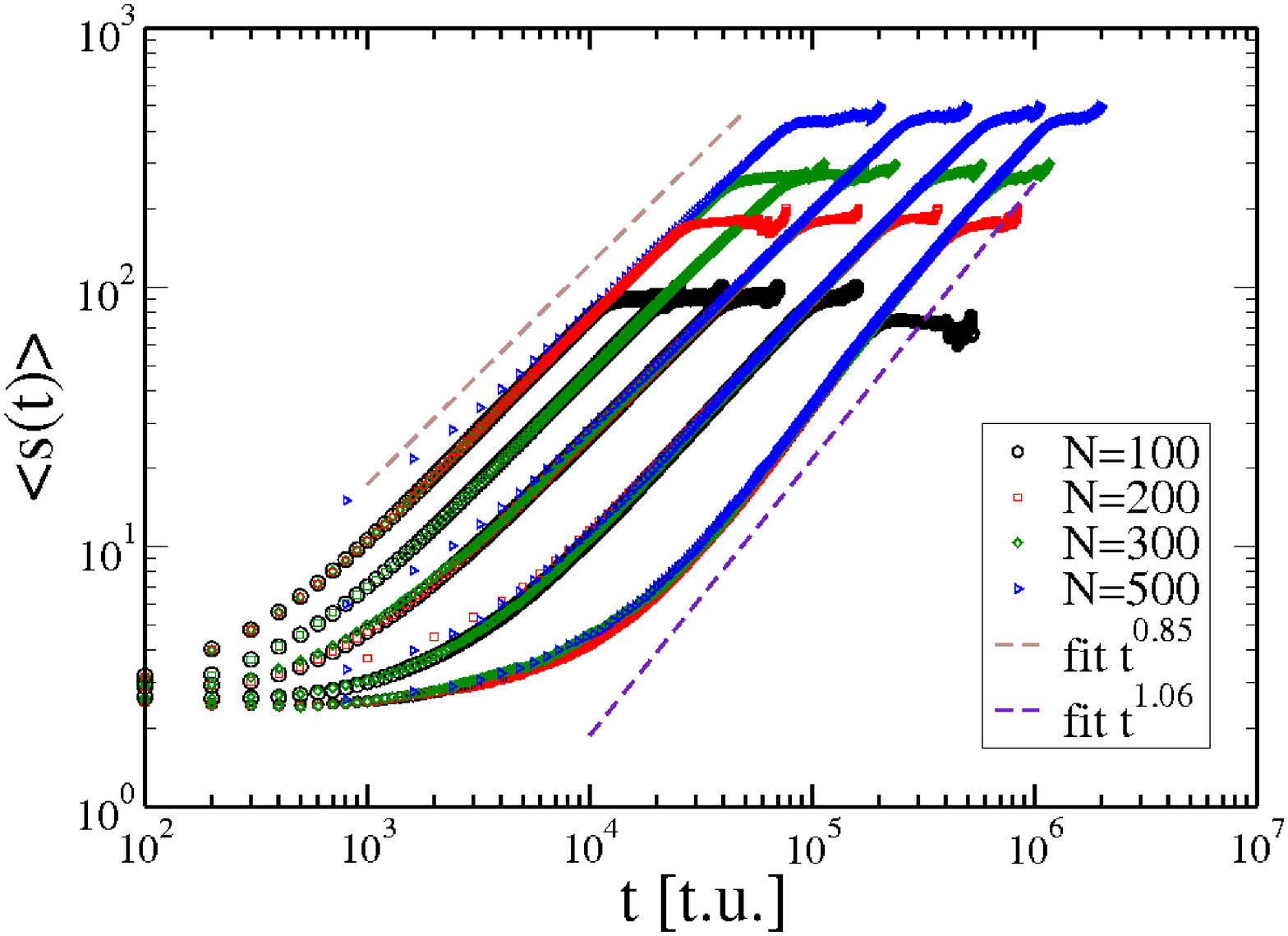}
\caption{(Color online) The average translocation coordinate $\langle s \rangle$ as function
of time for 4 different chains lengths $N = 100, 200, 300, 500$ and 5 different
forces $F=1, 2, 5, 10, 20$. The force increases from right to left and the chain
length increases from bottom to top.The translocation exponent $\beta$ decreases
for strong forces. }
\label{fig:savt}
\end{figure}

Much information about the translocation process can further be obtained by
examining the average value of the translocation coordinate $\langle
s(t)\rangle$. The results for $\langle s(t)\rangle$  at different forces and
chain lengths are  shown in Fig.~\ref{fig:savt}. It can be seen that for a fixed
force the behavior is largely independent of the chain length, that is, the
exponent $\beta$ in $\langle s(t)\rangle\simeq t^\beta$ does not depend on $N$.
Solely the plateau height which naturally (as for any Brownian motion on a
finite interval) marks the long time limit $\langle s(t \rightarrow
\infty)\rangle$ linearly depends on the chain length $N$.

One can also verify from Fig.~\ref{fig:savt} that the slope characterized by the
exponent $\beta$ becomes smaller with increasing strength of the driving force
$f$. Namely, $\beta \approx 1.06$ for small forces and $\beta \approx 0.85$ for
larger forces. These results could be compared with the corresponding
theoretical predictions given by Eq. (\ref{Trans_Monomers_Stat_2}) and Eq.
(\ref{Translocation_Law_Rouse}) (see also Fig.~\ref{Schematic}c ). This
basically reflects the same tendency which has already been seen in the $\langle
\tau \rangle \propto N^\alpha$ scaling relationship. Apparently, the
translocation time is defined as  $\langle s(\langle \tau \rangle) \rangle = N$,
i.e., $\langle \tau \rangle^\beta \sim N$ and we have $\beta \approx 1/\alpha$.
This relationships holds well in our MD-simulation. Therefore, the fact that
$\alpha$ grows with the force, which we have seen before, correlates with the
decrease of the exponent $\beta$.

\begin{figure}[ht]
\vspace*{1.0cm}
 \includegraphics[width=8.0cm]{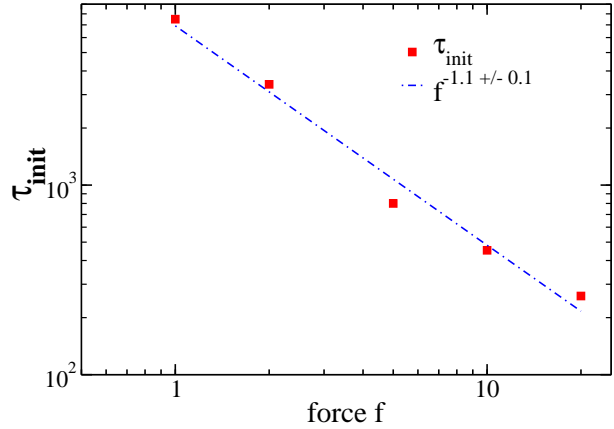}
\caption{(Color online) The time needed to create the first blob $\tau_{\rm init}$ as a
function of  force for three different chains lengths. It is clear that this
time is independent of the chain length and decreases approximately as
$\tau_{\rm init}\sim 1/f$. For large forces the force dependence  appears to
level off.}
\label{fig:fig5}
\end{figure}

From Fig.~\ref{fig:savt} it also easily inferred that during the  initial time
period $\langle s(t)\rangle$ remains to a very good approximation almost
constant. We identify this period with the blob initiation time (see Sec.
\ref{BI}), i.e., the time which is necessary to generate a first blob. In
Fig.~\ref{fig:savt} this characteristic time $\tau_{\rm init}$ manifests itself
as a first crossover from a constant value (within 10\%) to the scaling law
$\langle s(t) \rangle \sim t^\beta$. The time $\tau_{\rm init}$ is chain length
independent but inversely proportional to the  force $f$ (cf. Eq.
(\ref{Initiation_1})). This conclusion is supported by Fig.~\ref{fig:fig5}  where
the first crossover time $\tau_{\rm init}$ is plotted against the force.

\section{Conclusions}

In this work we derive scaling laws for the mean translocation time of a driven
polymer chain through a narrow pore, based on the ideas of tensile force
propagation along the polymer backbone \cite{Sakaue_1,Sakaue_2}. Our findings
can be summarized as follows:
\begin{enumerate}
 \item The translocation starts with the formation of initial Pincus blob (i.e.,
the first blob is generated immediately at the pore opening). The characteristic
time of the blob initiation is given by Eq.(\ref{Initiation_1}). Our
MD-simulation results essentially support the scaling prediction $\tau_{\rm
init} \sim 1/f$.

\item The initiation is followed by a tensile force transmission along the chain
backbone which is governed by the local balance of driving and drag forces. For
forces in the interval $N^{\nu} \ll af/k_BT < 1$ this leads to a trumpet regime
(see Fig. \ref{Transient}). The corresponding translocation time is given by Eq.
(\ref{Tau_Translocation}) where the first term, which corresponds to the force
transmission characteristic time, prevails  under the  condition Eq.
(\ref{Condition}). As a result, depending on the force strength and on chain
length (for Rouse dynamics), one expects a crossover from $N^{2\nu}/f^{1/\nu}$
to $N^{1+\nu}/f$, i.e., the translocation exponent $\alpha$ grows with
increasing force $f$ from $\alpha \approx 1.18$ to $\alpha \approx 1.59$.

One should note that the MD-simulation findings yield systematically smaller
values for the translocation exponent: $\alpha \approx 1.06$ and $\alpha \approx
1.33$ for weak and strong forces respectively. This is in agreement with other
simulation results \cite{Lehtola_1,Lehtola_2,Lehtola_3} and differs from
findings reported in ref. \cite{Luo_2}. The mentioned slight overestimation of
the translocation exponent $\alpha$ by the theory may be due to the role of
fluctuations which are not accounted for within the quasi-static approximation
used in this work. As for the force scaling, the theory predicts that the
scaling law ranges between $1/f^{1.66}$ and $1/f$ with the chain length
increasing. Again the MD-simulation gives smaller exponents: $1/f^{1.37}$ and
$1/f^{0.91}$ respectively.

\item Under the conditions, given by Eq. (\ref{Condition}), the number of
translocated segments $M(t)$ changes as $M(t)  = c_0 ( {\widetilde f}_a \:
{\widetilde t})^{1/(1 + \nu)}$ (see Eq. (\ref{Translocation_Law_Rouse})), i.e.,
in the scaling law $M(t)  \sim (f t)^\beta$ the exponent $\beta \approx 0.63$
for relatively large forces. Our MD-simulation result gives a slightly larger
value $\beta \approx 0.85$. The reason for that is the same as in the case of
the scaling law $\langle \tau \rangle \sim N^\alpha$: recall that $\beta \approx
1/\alpha$.

\item For strong forces the tensile force transmission follows either the
``stem-trumpet'' ($1 \ll af/k_BT \ll N^{\nu}$) or the ``stem'' ($N^{\nu} \ll
af/k_BT $) scenarios (see Fig. \ref{Transient}b, c). In both cases the
translocation time can be estimated as $\langle \tau \rangle = C_1 \: \tau_0
N^{1+\nu}/{\widetilde f}_a + C_2 \: \tau_0 \: N^{2 \nu}/{\widetilde f}_a $ where
the first term dominates under condition $N^{1 - \nu} \gg C_2/C_1$. In other
words, the translocation scaling exponent grows from $\alpha = 2\nu$ to $\alpha
= 1 + \nu$ as the chain length $N$ increases. This is in agreement with the
results of LD-simulation \cite{Luo_1} and also with the findings of Monte Carlo
investigations\cite{Slater}.

\end{enumerate} 

We believe that the present approach leads to a consequent and physically
plausible theory of driven translocation dynamics even though the theoretical
predictions regarding scaling coefficients do not fully agree with data from
computer experiments. The theory is based on the assumption that drag and
driving forces equal one another during the process of chain threading through
the pore.  Since at the present level of theoretical treatment fluctuations are
not taken into account, further developments of the theory should try to
incorporate fluctuations too. Also the  analytical model totally ignores
additional resistance forces which are imposed by the pore and by the
over-crowded beads at the pore exit (as it can be seen from
Fig.~\ref{DensityVSz}). These forces should be included in the general force
balance given by Eq.~(\ref{Balance}). On the other hand, these resistance forces
make the translocation dynamics even more sluggish, so that one can not explain
thus the observed decrease of the translocation exponents as compared to the
theoretical prediction. In principle, one could extend the present consideration
on the case of attractive pores where entering monomers may be captured and
reside for a while in the pore. Such studies have been carried out recently
\cite{Luo_3,Luo_4}. One should note, however, that this effect does not alter
the scaling laws for translocation even though the some additional details of
the process become thereby important. Evidently, more work is therefore needed
until a full understanding and entirely satisfactory description of
translocation dynamics is achieved. 
%The role of the pore becomes more demanding when the
%polymer-pore interaction potential has an attractive component, which probably
%simulates the difference between the $\alpha$-hemolysin channel and the
%synthetic nanopore \cite{Luo_3,Luo_4}.

\section{Acknowledgments}
 We should like to thank K.L. Sebastian and P. Rowghanian for fruitful
discussions. A.~Milchev thanks the Max-Planck Institute for Polymer Research in
Mainz, Germany, for hospitality during his visit in the institute. A.~Milchev
and V.~Rostiashvili acknowledge support from the Deutsche Forschungsgemeinschaft
(DFG), grant No. SFB 625/B4.

\begin{appendix}
\section{Integral and differential forms of the material conservation law}
\label{App}

Here we demonstrate why the calculation, performed in ref. \cite{Sakaue_2},
leads to a scaling result for $\tau_1$ which deviates from the one given by Eq.
(\ref{Tau_1}) in Section~\ref{TR}. In contrast to the {\em integral form} of the
material balance equation, given by Eq. (\ref{Material_Balance}) and used in the
present work, in ref. \cite{Sakaue_1,Sakaue_2} one uses its {\em differential}
form (cf. Eq. (5) in ref.~\cite{Sakaue_2}):
\begin{eqnarray}
 \left[\rho(x) \Sigma(x)\right]_{x= - X(t)} \: \left\lbrace \dfrac{d X(t)}{d t}
+  v(t) \right\rbrace = \dfrac{d N(t)}{d t}
\label{False}
\end{eqnarray}
As before, $\rho(x) \simeq g(x)/\xi(x)^3$ is the local density
and $\Sigma(x) \simeq \xi(x)^2 $ stands for the cross-sectional area so that
$\chi (x) = \rho(x) \Sigma(x) $ can be seen as the linear density.

One may readily see that the differential representation of the material
conservation, Eq.(\ref{False}), suffers from the problem that the linear density
$\left.\chi (x)\right|_{x=-X(t)}  = \left[\xi(x))/a]^{1/\nu -
1}/a\right|_{x=-X(t)}$ does not exist because it diverges! Indeed, it is evident
from Eq. (\ref{Solution_Blob_1}) that $\xi(x)$ at $x = - X(t)$ has an
integrable singularity. The physical reason for such behavior is clear: the
force at the free boundary is zero, therefore, the tensile blob size goes to
infinity (cf. Eq. (\ref{Zero}). Because of this, one should use from the very
beginning the integral form Eq. (\ref{Material_Balance}) and may turn to the
differential equation for $X(t)$ (cf. Eq. (\ref{Transmission_Diff})) only after
calculation of the involved integral.

Instead, Sakaue \cite{Sakaue_1,Sakaue_2} introduces a ``cut-off'' $x = - X(t) +
\xi_{X}(t)$, where $\xi_{X}(t)$ is the ``largest blob'' given by Eq.
(\ref{Blob_Max}). The implementation of the ``largest blob'' $\xi_{X}(t)$ as
well as the relationships given by Eq. (\ref{Border}) (at $X(t) \gg k_B T/f$)
and Eq. (\ref{Closure}) in Eq.(\ref{False}) leads eventually to a differential
equation for the front location $X(t)$ (in dimensionless units):
\begin{eqnarray}
 \tau_0 \left[  {\widetilde X}(t)^{\omega + 1}  {\widetilde f}_a^{\omega + 2 -
z} - c_0 {\widetilde X}(t) {\widetilde f}_a^{2 - z} \right] \dfrac{d {\widetilde
X}(t)}{d t} = 1 \label{Front_Location_Sacaue}
\end{eqnarray}
where ${\widetilde X}(t) = X(t)/a$, ${\widetilde f}_a = a f/k_B T$, 
$\omega = (1 - \nu)(z - 2)/ \nu(z - 1)$, and $c_0$ is a numerical coefficient.
The solution of this equation reads (cf. Eq. (\ref{Solution_Time}))
\begin{eqnarray}
 t = t_0 + \tau_0 {\widetilde f}_a^{\omega + 2 - z} {\widetilde X}(t)^{\omega + 2}
 \left[ 1 - c_0 /{\widetilde X}(t)^{\omega}{\widetilde f}_a^{\omega} \right]
\label{Solution_Time_Sakaue}
\end{eqnarray}
Finally, by equating  ${\widetilde X}(\tau_1) = N$,  Sakaue obtains the
characteristic time
\begin{eqnarray}
 \tau_1 \sim \tau_0 {\widetilde f}_a^{\omega + 2 - z } N^{(\omega + 2)\nu}
\label{Tau_Sakaue}
\end{eqnarray}
which is differs from our result, given by Eq. (\ref{Tau_1}). In the case of
Rouse dynamics ($z =2 + 1/\nu$), the scaling law Eq. (\ref{Tau_Sakaue}) leads to
$\alpha = (1 + \nu + 2\nu^2)/(1+ \nu) \approx 1.43$ and $\delta = 1/(1+\nu)
\approx 1.26$. Despite the fact that this value for the $\alpha$-exponent
correlates better with our MD-findings, one may doubt if the calculation based
on the ``cut-off'' trick  is trustworthy. Actually, this ``cut-off'' trick makes
the ``trumpet'' effectively shorter so that Sakaue's characteristic time $\tau_1
\sim N^{1.43}/f^{1.26}$   is  smaller than ours $\tau_1 \sim N^{1.59}/f$. 

Contrary to Sakaue's calculation, we use the {\em integral} form of the material
conservation law Eq.~(\ref{Material_Balance}) where the integral term can be
calculated {\em exactly} (without resorting to any ``cut-off'' trick). By
making use of the closure relation, Eq.~(\ref{Closure}), and the relationship
for the translocated monomers, $M(t)$ Eq.~(\ref{M_VS_Time}), we derive a
differential equation for ${\widetilde X}(t)$, Eq. (\ref{Transmission_Diff}),
which differs from the corresponding Eq.~(\ref{Front_Location_Sacaue}).

\end{appendix}

\end{document}